\documentclass[a4paper,11pt]{article}
\usepackage{jcappub} % for details on the use of the package, please see the JINST-author-manual
\usepackage{lineno}

\usepackage{newtxtext,newtxmath}
\usepackage[T1]{fontenc}

\DeclareRobustCommand{\VAN}[3]{#2}
\let\VANthebibliography\thebibliography
\def\thebibliography{\DeclareRobustCommand{\VAN}[3]{##3}\VANthebibliography}

\newcommand{\be}{\begin{eqnarray}}
\newcommand{\ee}{\end{eqnarray}}
\newcommand{\beq}{\begin{equation}}
\newcommand{\eeq}{\end{equation}}
\newcommand{\exclude}[1]{}

\usepackage{graphicx}	% Including figure files
\usepackage{amsmath}	% Advanced maths commands
\newcommand{\aqn}{{\rm AQN}}	% per cm-squared
\newcommand{\rmd}{{\rm d}}   % straight d for differential symbol

\title{\boldmath The Glow of Axion Quark Nugget Dark Matter: (I) Large Scale Structures}

\author[a]{Fereshteh Majidi,}
\author[a]{Xunyu Liang,}
\author[a]{Ludovic Van Waerbeke,}
\author[a]{Ariel Zhitnitsky,}
\author[a]{Michael Sekatchev,}
\author[b]{Julian S. Sommer,}
\author[b,c]{Klaus Dolag,}
\author[d]{Tiago Castro}
\affiliation[a]{Department of Physics and Astronomy,\\
University of British Columbia, \\
V6T 1Z1 Vancouver, BC, Canada}
\affiliation[b]{Universitäts-Sternwarte, Fakultät für Physik, \\
Ludwig-Maximilians Universität,\\
Scheinerstr. 1,\\
81679 München, Germany,}
\affiliation[c]{Max-Planck-Institut für Astrophysik,\\
Karl-Schwarzschild-Straße 1,\\
85741 Garching, Germany}
\affiliation[d]{Osservatorio Astronomico di Trieste, Villa Bazzoni \\
Via Bazzoni, 2, 34143 Trieste TS, Italy }

\emailAdd{fereshtehmajidi@phas.ubc.ca}

\abstract{Axion quark nuggets (AQN) are hypothetical, macroscopically large objects with a mass greater than a few grams and sub-micrometer size, formed during the quark-hadron %crossover 
transition. Originating from the axion field, they offer a possible resolution of the similarity between visible and dark components of the Universe, i.e. $\Omega_{\rm DM}\sim \Omega_{\rm visible}$ and observed matter-antimatter asymmetry. These composite objects behave as cold dark matter, interacting with ordinary matter and resulting in pervasive electromagnetic radiation throughout the Universe. This work aims to predict the electromagnetic signature in large-scale structures from this AQN-baryon interaction, accounting for thermal and non-thermal radiations.

We use \textit{Magneticum} hydrodynamical simulations to describe the realistic distribution and dynamics of gas and dark matter at cosmological scales. We construct a light cone encompassing a $1.4$ square degree area on the sky, extending up to redshift $z=5.4$, and we calculate the electromagnetic signature across a wide range of frequencies from radio, starting at $\nu \sim 1~\rm GHz$, up to a few keV X-ray energies. We find that the AQNs electromagnetic signature is characterized by global (monopole) and fluctuation signals. The amplitude of both signals strongly depends on the average nugget mass and the ionization level of the baryonic environment, allowing us to identify a most optimistic scenario and a minimal configuration. The signal of our most optimistic scenario is often near the sensitivity limit of existing instruments, such as FIRAS in the $\nu=[100-500]~\rm GHz$ range and the South Pole Telescope for high-resolution $\ell > 4000$ at $\nu = 95~\rm GHz$. Fluctuations in the Extra-galactic Background Light caused by the axion quark nuggets in the most optimistic scenario can also be tested with space-based imagers Euclid and James Webb Space Telescope. In general, our minimal configuration is still out of reach of existing instruments, but future experiments might be able to pose some constraints.

We conclude that the axion quark nuggets model represents a viable model for dark matter, which does not violate the canons of cosmology nor existing observations. A reanalysis of existing data sets could provide some evidence of axion quark nuggets if our most optimistic configuration is correct. The best chances for testing the model reside in 1) ultra-deep infrared and optical surveys, 2) future experiments to probe the frequency spectrum of the cosmic microwave background, and 3) low-frequency ($1~{\rm GHz} < \nu <100 ~\rm GHz$) and high-resolution ($\ell \gtrsim 10^4$) observations.
}

\begin{document}
\maketitle
\flushbottom

\section{Introduction}
The universe appears to be well described by a six-parameter model \cite{2020A&A...641A...6P}. The composition of matter has been precisely quantified, with 75\% attributed to Cold Dark Matter and 25\% to baryons. Despite this success, the physical nature of dark matter (DM) remains elusive. The challenge in identifying DM arises because the possible candidates span an immense range—90 orders of magnitude in mass and cross-section, from ultra-light bosons to primordial black holes that can weigh up to 10 solar masses \citep{2018Natur.562...51B}. This vast diversity makes the comprehensive probing of the parameter space unfeasible.

%\Red{Excess of electromagnetic radiation in the large scale structure provides a unique opportunity to constrain parameter space of dark matter models. We investigate a specific dark matter candidate, the axion quark nuggets (AQNs), which have a mass of grams and size of sub-micrometers \citep{Zhitnitsky:2002qa}. Unlike conventional dark matter candidates such as weakly interacting massive particles (WIMPs) and axions (e.g. review \citep{Marsh:2015xka} and references therein), AQNs interact strongly with baryonic gas. The AQN-induced electromagnetic radiation has a broad bandwidth from radio to X-ray.}
  
From a cosmological viewpoint, there is a fundamental difference between dark matter and ordinary matter. Ordinary matter structures started to grow after the matter-radiation decoupling, while dark matter, which is pressureless, started to form gravitational structures earlier, at the matter-radiation equality. Dark matter played a key role in structure formation by giving a head start to the gravitational collapse.
The key parameter describing dark matter as a pressureless fluid is the cross-section $\sigma$ to mass $M_{\rm DM}$ ratio, which must be sufficiently small: 
\be
\label{sigma/m}
\frac{\sigma_\textrm{DM-bar}}{M_{\rm DM}}\ll  1~{\rm cm^2\, g^{-1}}.
\ee

The observational constraints on the DM-baryon scattering are from Big Bang nucleosynthesis \cite{Mohapatra:1998nd,Cyburt:2002uw}, spectral distortion of the Cosmic Microwave Background (CMB) \cite{Ali-Haimoud:2015pwa,Ali-Haimoud:2021lka}, the temperature and polarization anisotropies of the CMB \cite{Dubovsky:2001tr,Chen:2002yh,Dvorkin:2013cea,Gluscevic:2017ywp,Boddy:2018kfv,Boddy:2018wzy,Xu:2018efh,Li:2018zdm,Li:2022mdj}, the thermal history of the intergalactic medium \cite{Tashiro:2014tsa,Munoz:2015bca,Munoz:2017qpy}, the Lyman-$\alpha$ forest \cite{Chen:2002yh,Dvorkin:2013cea,Xu:2018efh,Rogers:2021byl}, Milky Way satellite galaxies \cite{Nadler:2019zrb,DES:2020fxi,Maamari:2020aqz}\footnote{For Ref. \cite{Nadler:2019zrb}, there is an erratum [Astrophys. J. Lett. 897, L46 (2020)]}, and the gas heating in galaxies \cite{Chivukula:1989cc,Bhoonah:2018wmw,Wadekar:2019mpc,Bhoonah:2020dzs} and galaxy clusters \cite{Qin:2001hh,Chuzhoy:2004bc,Hu:2007ai}. 
%\Red{A bit confused by this sentence, is there a verb missing? XL: yes. missing ``are"}. 
Generally, the constraints suggest $\sigma_{\rm DM-bar}/M_{\rm DM} \lesssim 10^{-3}-0.1{\rm\,cm^2\, g^{-1}}$ for the elastic DM-baryon scattering. However, the specific value greatly depends on the characteristics of the dark matter candidate. These are not very strong constraints compared to the cross-section for Weakly Interacting Massive Particles (WIMPs) $\sigma_{\rm WIMP-bar}/M_{\rm WIMP} \lesssim 3.4\times10^{-25}{\rm\,cm^2\, g^{-1}}$ for a typical mass $M_{\rm WIMP}\in(10,10^3) {\rm\,GeV}$ from recent direct detection \cite{PandaX:2022xqx,LZ:2022lsv,XENON:2023cxc}. For light DM of mass in the range of $(10^{-4},0.1){\rm\,GeV}$, the cross-section above $\sim10^{-30}{\rm\,cm^2}$ is excluded by local observation and neutron stars \cite{Cyburt:2002uw,Bringmann:2018cvk,Bramante:2021dyx,Bramante:2022pmn}.  For even lighter particles such as axions, the axion-nucleon cross section is less than $\sim10^{-44}{\rm\,cm^2}$ from neutron stars and supernova 1987A; see e.g. recent studies \cite{DiLuzio:2021qct,DiLuzio:2021ysg,Carenza:2023wsm} and references therein. If DM is a new elementary particle, it has to be lighter than a few hundred TeV due to the unitarity limits \cite{Griest:1989wd}, and {\bf the condition \eqref{sigma/m} imposes an upper-bound on the cross-section $\bf \sigma \ll 10^{-19}~\rm cm^2$}. Detailed information on cosmological constraints, direct detection, and properties of DM can be found in the recent review \cite{Cirelli:2024ssz}.

In the present work we consider a fundamentally different type of DM where the condition (\ref{sigma/m}) is satisfied {\bf with a cross-section exceeding the upper-bound of a new elementary particle hypothesis. This is possible if dark matter is a composite, macroscopically large object with a large mass.} We consider a specific generalization of the Witten's quark nuggets \citep{1984PhRvD..30..272W,PhysRevD.30.2379,1984Natur.312..734D}, the so-called axion quark nuggets (AQNs) \cite{Zhitnitsky:2002qa} {\bf where DM is typically $M_{\rm DM}\gtrsim 1~\rm g$, nuclear density with a size of sub-micrometer}. Unlike conventional dark matter candidates such as WIMPs and axions (e.g. review \citep{Marsh:2015xka} and references therein), AQNs interact strongly with baryonic matter.
The reason is that AQNs form during the quark-hadron transition, when matter and antimatter have not yet annihilated, therefore some AQNs are made of antimatter which leads to broad bandwidth, from radio to X-ray, electromagnetic radiation when AQNs collide with normal matter.

Motivated by the unique emission properties of AQNs, we investigate the AQN signature from large-scale structures from the present-day up to redshift $z=5.34$ using the \textit{Magneticum} hydro simulations. We make predictions for a wide bandwidth range, from radio, where the CMB dominates, to infrared, optical and X-ray and investigate both the global (monopole) intensity and anisotropies. In general, we find that the AQN signal takes the form of radiation excess or anomalies, correlated with large scale structures. We discuss the potential detectability of the signal with existing and future observatories.

The paper is structured as follows: In section \ref{sec:AQNmodel} we briefly review the AQN formation mechanism. In section \ref{sec:emission_mec} we describe the AQN emission mechanisms and the role of the baryonic environment. In section \ref{subsec:datacubes} we describe the simulations used in this work and section \ref{subsec:Computing I_nu} shows how the light cone is constructed and how the AQN signal is calculated from the simulation. We will discuss the results in section \ref{sec:results}, and in section \ref{sec:conclusion}, after summarizing our main results, we discuss possible future work, including a description of existing data that could yield a detection and long term theoretical developments.

\section{The Axion Quark Nugget model}
\label{sec:AQNmodel}

The model was initially proposed by \cite{Zhitnitsky:2002qa} as an explanation for why the dark matter (DM) mass density and the visible matter mass density are of the same order of magnitude, denoted as $\Omega_{\rm DM}\sim \Omega_{\rm visible}$. The Axion Quark Nugget (AQN) construction shares many similarities with Witten's quark nuggets, as referenced in \citep{1984PhRvD..30..272W}. This form of DM is "cosmologically dark," meaning that its interaction cross-section to mass ratio, $\sigma/m_{\rm DM}$, is significantly less than $1~{\rm cm^2g^{-1}}$. This numerically small ratio reduces many electromagnetic signatures that would typically be associated with a strongly-interacting DM candidate, thereby ensuring that AQN kinematics align with those of cold dark matter (CDM), which is pressureless. In the subsequent sections, we will briefly review the primary steps involved in the formation of an AQN. For more detailed information, interested readers are referred to a recent concise review \cite{Zhitnitsky:2021iwg}.

\subsection{The quark-hadron transition and the AQN formation}

 The formation of AQNs is dependent on the so-called $N=1$ QCD axion domain wall bubbles, as detailed in references \citep{Liang:2016tqc,Ge:2017ttc,Ge:2017idw,Ge:2019voa}. It is generally assumed that these closed domain wall bubbles are abundantly produced and collapse without leaving observable traces. Within the AQN model, however, some of these bubbles manage to persist by accreting quarks, thereby forming AQNs in which the internal Fermi pressure counterbalances the external wall pressure. The accretion process becomes particularly effective during the quark-hadron transition when the temperature of the universe reaches approximately $\sim 170~{\rm MeV}$\footnote{This energy scale corresponds to the Quantum ChromoDynamic (QCD) energy scale.}. The quarks confined within these bubbles then exhibit a pronounced matter-antimatter asymmetry due to the charge separation, a characteristic effect of the ${\cal{CP}}$-odd axion field's interaction with matter and antimatter.
 As matter-antimatter annihilation persists in the plasma, both matter and antimatter nuggets form in proportions comparable to that of the newly created hadronic matter (protons and neutrons) outside the closed bubbles.

As the system continues to cool down, baryons and anti-baryons outside the bubbles persist in annihilating each other until all anti-baryons are depleted. The nuggets formed during the QCD transition may consist of both {\it matter} and {\it antimatter} AQNs. A significant implication of this characteristic—considering the Universe's total baryon charge is zero—is that the densities of dark matter (DM), $\Omega_{\rm DM}$ (represented in this context by both matter and antimatter nuggets), and visible baryonic matter, $\Omega_{\rm visible}$, will naturally align to the same order of magnitude, $\Omega_{\rm DM} \sim \Omega_{\rm visible}$. This correlation occurs independently of specific model details, such as the axion mass $m_a$. {\bf The relative proportions between antimatter nuggets, matter nuggets and baryons differ by order one, but the exact calculation cannot be done without a full solution to the strongly coupled QCD, currently not achievable. We therefore adopt the proportions (3:5/2:5/1:5) as a benchmark, as suggested in \cite{Zhitnitsky:2021iwg}.}
%\footnote{An exact calculation of the mass densities would require a full solution to the strongly coupled QCD problem, which is currently unachievable.}.

\subsection{Stability of AQNs}

The interior of nuggets is composed of matter in the form of a diquark condensate, which exists in the color superconducting phase. This phase is recognized as the lowest energy state under sufficiently high pressures and may be present in the cores of neutron stars (NS),
%Additionally, it is noted that the mass density in this phase is slightly greater than nuclear density, 
as discussed in the review \cite{Alford:2007xm}. A practical approach to understanding nuggets is to consider them analogous to miniature NS, hold together by the axion domain wall tension, with masses ranging from a gram to several tons. Importantly, the energy per baryon charge in the quark-matter nuggets is lower than in hadrons, indicating that AQNs are more stable than protons and neutrons.

There are several additional elements in the AQN model in comparison with the older constructions of the Witten's nuggets \citep{1984PhRvD..30..272W, PhysRevD.30.2379, 1984Natur.312..734D}. Firstly, AQNs can be made of {\it matter} as well as {\it antimatter} during the QCD transition as we already mentioned. Secondly, there is an additional stabilization factor for the nuggets provided by the axion domain walls which  are copiously produced  during the   QCD  transition. This additional element helps to alleviate a number of  problems with the Witten's original model. In particular, a first-order phase transition (which is now known to not occur in QCD) was a required feature for the Witten's nuggets to be formed.  It is not required for AQN formation because the axion domain wall (with internal QCD substructure) plays the role of the squeezer, keeping the quark (anti-quark) matter stable.
Another problem of the old construction \citep{1984PhRvD..30..272W, PhysRevD.30.2379, 1984Natur.312..734D} is that nuggets likely evaporate on the Hubble timescale. For the AQN model, this is not the case because the vacuum-ground-state energies inside (the colour- superconducting phase) and outside the nugget (the hadronic phase) are drastically different. Therefore, these two systems can coexist only in the presence of an external pressure, provided by the axion domain wall, which is an inevitable feature of the AQN construction. This should  be contrasted with the original model \citep{1984PhRvD..30..272W, PhysRevD.30.2379, 1984Natur.312..734D},  which is assumed to be stable at zero external pressure. These differences have important phenomenological implications:
In Witten's model, the nuggets were postulated to be stable even at zero external pressure, as mentioned above. In scenarios where they collide with a neutron star, the model suggests that the entire neutron star would be converted into a quark star. In the AQN model
%axion domain walls are abundantly produced during the QCD transition, addressing several issues inherent in the original Witten's model.
a matter-type AQN will not transform an entire neutron star into a quark star. This is because the quark matter within AQNs is maintained by the pressure from external axion domain walls, limiting its expansion to only a small region rather than the entire star.  

\subsection{Survival of AQNs and mass function}\label{sect:mass-distribution}

\begin{table}%[h!]
	\centering
    \caption{Mass $m_{\rm AQN}$ and radius $R_{\rm AQN}$ of an AQN using the nuclear mass density $\rho_{\rm nucl}=3.5\times10^{14}{\rm\,g\,cm^{-3}}$ \citep{Raza:2018gpb} for the quark-gluon material. To calculate the number density $n_{\rm AQN}$, we use a dark matter mass density $\rho_{\rm DM}=0.3~{\rm GeV~m^{-3}}$, which corresponds to the average cosmological DM mass density today ($z=0$) for a $\Omega_{\rm  DM}=0.266 $ universe, and assuming that all dark matter is made of AQNs. The number density is $n_{\rm AQN}=\frac{2}{3}\times\frac{3}{5}\frac{\rho_{\rm DM}}{m_{\rm AQN}}$, which considers only the antimatter AQNs with axion contribution excluded.}
	\begin{tabular}{l c c l} 
		\hline\hline
        $m_{\rm AQN}\,[{\rm g}]$ & $R_{\rm AQN}\,[{\rm cm}]$ & $n_{\rm AQN}\,[{\rm m^{-3}}]$ \\ 
		\hline
  		1 & $8.8\times10^{-6}$  & $9.0\times 10^{-25}$  \\ 
		10 & $1.9\times 10^{-5}$  & $9.0\times 10^{-26}$   \\ 
		100 & $4.1\times 10^{-5}$ & $9.0\times 10^{-27}$  \\ 
		1000 & $8.8\times10^{-5}$ & $9.0\times 10^{-28}$ \\ 
		\hline\hline
	\end{tabular}
	\label{tab:Some AQN quantities}
\end{table}

The mass $m_{\rm AQN}$ of an AQN is the only fundamental parameter of the model, though it is relatively insensitive to the fundamental mass of the axion field $m_a$. For a given $m_a$, the AQN size cannot exceed $R_{\rm AQN}^{\rm max} \sim m_a^{-1}$ \cite{Liang:2016tqc}. Therefore, the AQN mass follows a distribution with a high mass cutoff constrained by the axion mass $m_a$ which can only be determined experimentally. Various observational limits on Macroscopic Dark Matter size imposes a constraint on their number density, which, in the context of the AQN model, allows the nugget mass to be anywhere between $1~{\rm g}$ and higher. According to \cite{Lawson:2019cvy}, the most stringent constraint on $\langle m_{\rm AQN}\rangle$ is greater than 5 grams, derived from the lack of detection at the IceCube observatory. Similar constraints are reported by the ANITA experiment and from geothermal data \citep{Gorham:2012hy}. While the exact mass distribution function of AQNs remains unknown, percolation simulations suggest that it follows a power-law distribution \citep{Ge:2019voa}. 

Table \ref{tab:Some AQN quantities} presents the sizes and number densities of AQNs for various masses. Given their nuclear density, these objects are extremely compact, and their substantial mass, relative to typical particle physics standards, results in a notably low number density\footnote{For comparison, the average number density of WIMPs in the universe today is approximately $n_{\rm WIMP}\approx 10^{-3}{\rm m^{-3}}$ for a WIMP mass of $100-200{\rm GeV}$, which is at least 20 orders of magnitude greater than $n_\aqn$}. It is sometimes beneficial to utilize the baryon charge $B$ (equivalent to the proton mass) as a proxy for mass \citep{Lawson:2019cvy}:
\begin{equation}
B \gtrsim 3\times 10^{24}\,.
\end{equation}
The AQN mass, geometrical cross-section $\sigma_{\rm geo}$ and number density can be written as \citep{Zhitnitsky:2021iwg}:
\begin{align}
&m_{\rm AQN} \approx 16.7\,\left(\frac{B}{10^{25}}\right)~ {\rm g}\nonumber \\
&\sigma_{\rm geo} \equiv \pi R_\aqn^2 \approx  1.59\times10^{-9}\left(\frac{R_{\rm AQN}}{2.25\times10^{-5}{\rm\,cm}}\right)^2~{\rm\,cm^2}\nonumber\\
&n_{\rm AQN} \equiv \frac{\rho_{\rm DM}}{m_\aqn}  \sim  0.3\times 10^{-25}  \left(\frac{10^{25}}{B}\right)~{\rm cm^{-3}}
\end{align}
where we have used $\rho_{\rm DM}=0.3~\rm GeV~cm^{-3}$.
Antimatter AQNs experience mass loss, denoted as $\rm \Delta B$, even after their formation, primarily due to interactions and subsequent annihilations with baryons. It has been calculated that the ratio $\Delta B/B\ll 1$ remains consistent across all cosmic epochs, including during the Big Bang Nucleosynthesis, as well as the pre- and post-recombination and galaxy formation periods \citep{Flambaum:2018ohm, Zhitnitsky:2006vt, Lawson:2018qkc, Ge:2019voa}. This finding is further corroborated by independent analyses from \cite{Santillan:2020lbj} and \cite{SinghSidhu:2020cxw}.

For the remainder of this paper, our focus will be on the electromagnetic signatures of antimatter nuggets. We will consider a fixed nugget mass $m_{\rm AQN}$, with consistent nuclear density and radius, denoted as $n_{\rm AQN}$ and $R_{\rm AQN}$, respectively. Although our calculations can be extended to account for a mass distribution, we opt for fixed values for simplicity. The relaxing of this constraint will make the allowed window for $B$ broader.

\section{The Emission Mechanism}
\label{sec:emission_mec}

\subsection{The Energy Budget}
\label{sec:energy_budget}

In the AQN model, matter-based AQNs are surrounded by a cloud of electrons, while antimatter AQNs are surrounded by a cloud of positrons, both configurations serving to maintain electrical neutrality. For the purposes of this paper, the cloud surrounding antimatter AQNs will be referred to as the electrosphere. When a matter AQN collides with conventional matter, the energy exchange is minimal, leading to negligible radiation. Conversely, collisions between antimatter AQNs and regular matter (such as protons, hydrogen, and helium) results in annihilation events that release substantial amounts of energy. Within the AQN framework, the absence of an electromagnetic signature from dark matter is not due to exceedingly weak interactions with baryons\footnote{as it should be the case for WIMPs.}, but it is instead attributed to the significantly lower number density of AQNs compared to baryons ($n_{\rm AQN} \ll n_{\rm b}$), despite the strong interaction  between antimatter AQNs and baryons\footnote{Note that "strong interaction" refers to the large dark matter-baryon interaction cross-section, not to the strong nuclear force.}.

The collision rate of an AQN with the surrounding baryonic matter is given by:
\begin{equation}
\Gamma = \sigma_{\rm eff} n_{\rm b}\Delta {\rm v}\,,
\label{eq:Gamma}
\end{equation}
where $\sigma_{\rm eff}$ is the effective cross-section of the AQN, $n_{\rm b}$ is the baryon number density, and $\Delta {\rm v}=|\vec{\rm v}_{\rm AQN}-\vec{\rm v}_{\rm b}|$ is the relative speed between an AQN and a baryon. The effective cross-section $\sigma_{\rm eff}$ describes the area around the AQN capable of capturing baryonic matter, potentially triggering an annihilation event. $\sigma_{\rm eff}$ is not necessarily equal to the geometrical cross-section $\sigma_{\rm geo}$. This is the case when the AQN acquires a non-zero electrical charge and the baryonic environment is ionized, and the Coulomb force could significantly change the cross-section. The calculation of $\sigma_{\rm eff}$ under this condition will be explored in Section \ref{sigma_eff}.

Nuggets predominantly collide with atomic or ionized hydrogen and helium, the most abundant elements in the universe. For each collision, the incoming baryonic matter will either rebound due to quantum reflection \citep{Forbes:2006ba} or be captured, penetrate the nugget's core, and annihilate. The likelihood of quantum reflection has a probability $(1-f)$. The specific value of $f$ depends on the model, but assuming $1-f \sim 0.9$ is considered reasonable \citep{Forbes:2006ba}. For matter that penetrates the core, the energy available for photon production from the annihilation is twice the rest mass of the incoming particle\footnote{Strictly speaking, axions contribute a significant fraction of the radiation, approximately 1/3 of the total \citep{Ge:2017idw}. We adjust for the axion contribution by defining the number density:

\begin{equation*}
n_{\rm AQN}
=\frac{2}{3}\times\frac{3}{5}\frac{\rho_{\rm DM}}{m_{\rm AQN}}\,,
\end{equation*}
where $2/3$ is the non-axion fraction of a nugget mass, $3/5$ is the fraction of antimatter nuggets as explained in the preceding subsection.
}, and the kinetic energy is negligible.
A fraction $g$ of this energy will be emitted as a short pulse of non-thermal photons at the point of impact, while a remaining fraction $(1-g)$ will be transfered as heat into the AQN anti-quarks core and subsequently to the electrosphere which will emit thermally\footnote{Similar to $f$, the factor $g$ is also highly model dependent and must be less than unity. We use $g \simeq 0.1$, as discussed in \cite{Forbes:2008uf}. A small fraction of the annihilation energy will convert to subatomic particles, which we are neglecting.}. As will be discussed in Section \ref{cooling_time}, the substantial heat capacity of the anti-quark matter results in a cooling time significantly longer than the average time between collisions. Consequently, the AQN reaches thermal equilibrium at some temperature $T_\aqn$, and the electrosphere emits thermally. Meanwhile, the energy released at the point of impact can escape the system immediately via non-thermal processes.

For simplicity, we will assume that the incoming particle is a proton, thus corresponding to $2~\rm GeV$ of energy. The energy $\Delta E_{\rm ann}^{\rm th}$ emitted thermally is therefore given by:

\begin{equation}
    \Delta E_{\rm ann}^{\rm th} = 2{\rm\,GeV}f(1-g)\,,
\end{equation}
which corresponds to the power: %\Red{Two ways of differential symbols used in this paper: $d$ and ${\rm d}$. Please pick one, and I would replace all of them.}

\begin{equation}
    \frac{\rmd E_{\rm ann}^{\rm th}}{\rmd t} = \Delta E_{\rm ann}^{\rm th} \Gamma = 2{\rm\,GeV}f(1-g)\,\sigma_{\rm eff}\,n_{\rm b}\,\Delta{\rm v}\,.
    \label{eq:E_ann}
\end{equation}

The non-thermal photons will be radiated away in a short pulse during the annihilation event, delivering an average power (per event): %\Red{I replaced all quantities with the form ``$X^{\rm non-th}$'' in the paper with ``$X^\textrm{non-th}$''. The hyphen is shorter and looks nicer}

\begin{equation}
    \frac{\rmd E_{\rm ann}^\textrm{non-th}}{\rmd t} = 2~{\rm GeV}fg\sigma_{\rm eff}~n_{\rm b}~\Delta{\rm v}\,.
    \label{eq:E_Xray}
\end{equation}

Therefore, any annihilation event will produce two outcomes: 1- it sustains the electrosphere at an equilibrium temperature $T_\aqn$, which consequently emits thermal radiation continuously, and 2- it generates a pulse of non-thermal photons at the point of impact.

\subsection{The Thermal Emission}
\label{sec:thermal}
\subsubsection{Radiated Power}

We now focus on the electromagnetic radiation emitted from the heated electrosphere. The primary factor influencing photon emissivity in this scenario is the nugget temperature, $T_{\rm AQN}$. As discussed in Section \ref{sec:energy_budget}, $T_{\rm AQN}$ is determined by the annihilation rate within the nugget, which itself depends on the surrounding baryonic environment (whether it is ionized or not) and the collision rate.

When the anti-quark core of the nugget is at zero temperature ($T_{\rm AQN} = 0$), all positrons occupy Fermi levels beneath the surface of the nugget, meaning that they behave similarly to degenerate electrons in a white dwarf. Owing to the Pauli exclusion principle, positrons in this state cannot change their quantum states, and thus, no radiation is produced. However, at non-zero temperatures, a thin layer of positrons forms above the core, occupying Boltzmann states, which enables photon emission. As shown in \cite{Forbes:2009wg}, positrons near the core are relativistic, while those further from the core are not, with the latter emitting thermal photons through Bremsstrahlung \cite{Forbes:2006ba, Forbes:2008uf}. The flux density emitted by a single nugget, $ F_{\rm tot}^{\rm th}(T_\aqn)$, is derived in \citep{Forbes:2008uf}:

\begin{equation}
F_{\rm tot}^{\rm th}(T_\aqn) = \frac{16\alpha^{5/2}}{3\pi}T_\aqn^4\left(\frac{T_\aqn}{m_e}\right)^{1/4}\,,
\label{eq:dFtot}
\end{equation}
where $\alpha$ is the fine structure constant, and $m_e$ the electron mass. In this work, we have adopted natural units, setting $\hbar$, $k_B$, and $c$ to 1. The total power emitted by the electrosphere is isotropic and can be formulated as \footnote{In C.G.S. physical units, the quantity $ F_{\rm tot}$ is measured in $\rm erg~s^{-1}cm^{-2}$. While we use natural units in the main text to streamline discussions involving orders of magnitude. For those accustomed to the C.G.S. system, refer to Table \ref{tab:Eqs_cgs2nat} in Appendix \ref{app: units}.}:

\begin{equation}
    \frac{\rmd E_{\rm ann}^{\rm th}}{\rmd t}=4\pi R_{\rm AQN}^2 F_{\rm tot}^{\rm th}(T_\aqn)\,.
    \label{eq:equi_condition}
\end{equation}

It is important to note that Eq. \eqref{eq:dFtot} and \eqref{eq:equi_condition} are only valid if the AQN is in thermal equilibrium. The equilibrium temperature of the electrosphere, $T_\aqn$, is determined by the annihilation rate inside the core, which in turn depends on the effective cross-section $\sigma_{\rm eff}$, as indicated by Eq. \eqref{eq:Gamma}, and on the cooling time. In the following section, we will calculate $\sigma_{\rm eff}$, and in Section \ref{cooling_time}, we will determine the cooling time. A cooling time that exceeds the collision time is necessary to maintain the thermal equilibrium of the AQN electrosphere.

\subsubsection{Effective cross-section $\sigma_{\rm eff}$}
\label{sigma_eff}

To calculate $\sigma_{\rm eff}$, it is essential to consider that the AQN may acquire a net negative electrical charge $Q$ at non-zero temperatures. This occurs because, when $T_\aqn > 0$, the least bound positrons can escape the system, leading to $Q<0$. This will increase $\sigma_{\rm eff}$ when the surrounding baryonic environment contains positively charged ions.

The charge $Q$ is calculated under the assumption that positrons with kinetic energy exceeding the potential energy will escape. This relationship is formalized as follows \cite{Zhitnitsky:2023znn}:

\begin{equation}
|Q| \simeq 4\pi R_{\rm AQN}^2 \int_0^\infty n(z,T_{\rm AQN}) \rmd z = \frac{4\pi R_{\rm AQN}^2}{\sqrt{2\pi\alpha}} (m_e T_{\rm AQN}) \left(\frac{T_{\rm AQN}}{m_e}\right)^{1/4}\,.
\label{eq:Q}
\end{equation}
Here, $n(z,T_{\rm AQN})$ represents the local density of positrons at a distance $z$ from the nugget’s surface, calculated using the mean field approximation \cite{Forbes:2008uf}:
\begin{equation}
n(z,T_{\rm AQN}) = \frac{T_{\rm AQN}}{2\pi\alpha} \frac{1}{(z+\bar z)^2}\,.
\end{equation}
In this equation, $\bar z$ is an integration constant determined numerically to align with the Boltzmann regime.

\begin{equation}
\bar z^{-1}\simeq \sqrt{2\pi\alpha} m_e\left(\frac{T_\aqn}{m_e}\right)^{1/4}\,.
\end{equation}
When $Q<0$, the AQN will capture surrounding free protons (or any positively charged ionized nuclei) with an effective impact parameter $R_{\rm eff}$ larger than the AQN radius $R_\aqn$. The equality between the kinetic energy and the Coulomb potential energy is used to define $R_{\rm eff}$:

\begin{equation}
    \frac{\alpha Q}{R_{\rm eff}}\sim T_{\rm gas}\,,
    \label{eq:coulomb}
\end{equation}
where $T_{\rm gas}$ denotes the temperature of the surrounding ionized gas. The interpretation of Eq. \eqref{eq:coulomb} is straightforward: with a constant charge $Q$, a lower $T_{\rm gas}$ leads to charged particles being more easily deflected by the Coulomb potential of the AQN, thus increasing the effective radius, $R_{\rm eff}$. Conversely, a higher $T_{\rm gas}$ reduces $R_{\rm eff}$.

\begin{figure}
    \begin{center}
	\includegraphics[width=13cm]{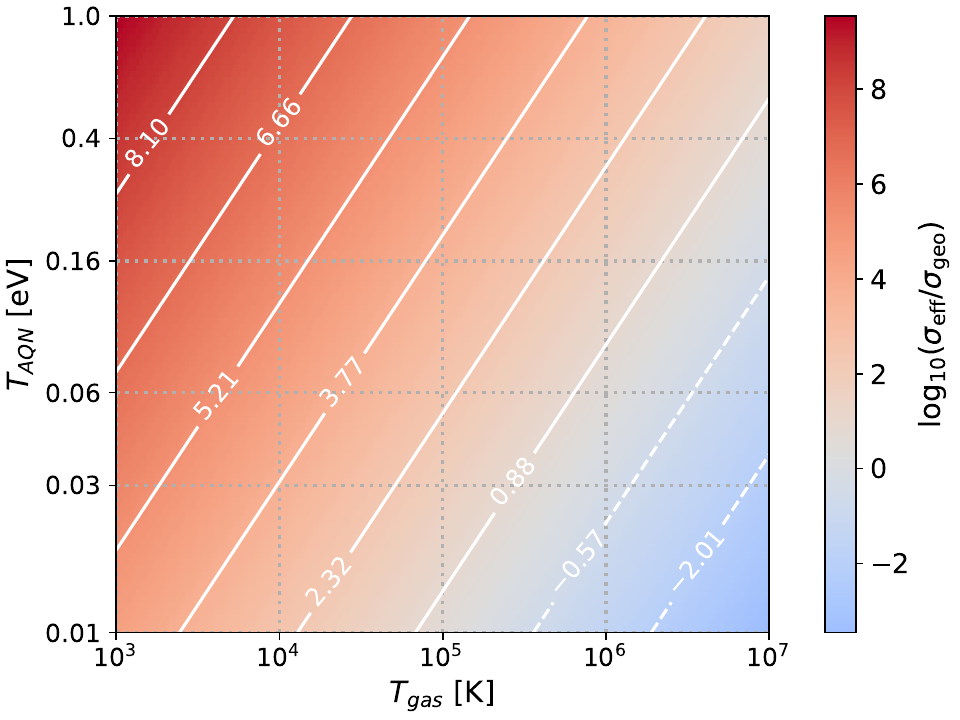}
	\end{center}
    \caption{Cross-section ratio $\bf\rm\log_{10}(\sigma_{\rm eff}/\sigma_{\rm geo})$ given by Eq. (\ref{eq:Reff}) for an AQN mass $m_\aqn=10{\rm\,g}$ and with $\Delta \rm v=10^{-3}c$. The gradian shows how the effective cross-section $\sigma_{\rm eff}$ compares to the geometrical cross-section $\sigma_{\rm geo}$ when the baryons temperature $T_{\rm gas}$ of the environment around the AQN varies for a given AQN temperature $T_\aqn$. {\bf The dashed line indicate the region where $\sigma_{\rm eff} < \sigma_{\rm geo}$ according to Eq. (\ref{eq:Reff})}. It is expected that, in real physical environments, one has to set $\sigma_{\rm eff}=\sigma_{\rm geo}$ {\bf in this region because the hot baryons move too fast to be captured by the charged AQN, unless they are on a head-on collision course with the AQN} (see conditions 1 in Section \ref{sigma_eff}).}
    \label{fig:cross_section_ratio}
\end{figure}

Combining Eqs. \eqref{eq:Q} and \eqref{eq:coulomb} and defining $\sigma_{\rm eff} = \pi R_{\rm eff}^2$, we obtain:

\begin{equation}
\frac{\sigma_{\rm eff}}{\sigma_{\rm geo}} = \left(\frac{R_{\rm eff}}{R_{\rm AQN}}\right)^2 = 8\alpha m_e^2 R_{\rm AQN}^2 \left(\frac{T_{\rm AQN}}{T_{\rm gas}}\right)^2\sqrt{\frac{T_{\rm AQN}}{m_e}}\,.
\label{eq:Reff}
\end{equation}

It is very important, for the rest of this paper, to emphasize that there are two limiting cases where Eq. (\ref{eq:Reff}) should not be used:

\bigskip

Condition 1- When $T_{\rm gas}$ is high enough to result in $R_{\rm eff} < R_\aqn$, the baryons in the AQN rest frame move so rapidly that their capture becomes unlikely, and the cross-section effectively reduces to the geometrical cross-section, i.e. head-on collisions\footnote{Note that the only situation in which $R_{\rm eff} < R_{\rm AQN}$ physically occurs is when the baryonic matter is negatively charged, a possibility not considered in this work.}.

Condition 2- When the surrounding gas is not ionized. This typically occurs when $T_{\rm gas}$ falls below a certain threshold. However, in general, the specific conditions for non-ionized gas depend on the particular astrophysical environment, and a simple temperature threshold is not a sufficient criterion for deciding when the gas is ionized or not.

\bigskip

If either condition 1 or 2 applies, the geometrical cross-section should be used, i.e., $\sigma_{\rm eff} = \sigma_{\rm geo}$. It is crucial to recognize that when neither of these conditions is met, the quantum reflection probability $1-f$, as defined in Section \ref{sec:emission_mec}, effectively becomes zero. This occurs because, once captured, the baryon is certain to annihilate with the AQN anti-quark matter ($f=1$), although with a time delay that is irrelevant for our purpose.

Using Eq. \eqref{eq:Reff}, $\sigma_{\rm eff}/\sigma_{\rm geo}$ is shown in Figure \ref{fig:cross_section_ratio} as a function of $T_{\rm gas}$ and $T_\aqn$ for an AQN with a mass of $m_{\rm AQN} = 10{\rm g}$\footnote{The range of $T_{\rm gas} \sim 10^3-10^4{\rm K}$ typically corresponds to cosmic voids, while a $T_{\rm gas}$ exceeding $10^6{~\rm K}$ characterizes galaxy clusters.}. The bottom-right corner of Figure \ref{fig:cross_section_ratio}, where $\sigma_{\rm eff} < \sigma_{\rm geo}$, indicates where condition 1 is applicable. Conversely, on the left side of the figure, where $T_{\rm gas}$ is lower, the baryonic matter is likely to have recombined into  neutral atomic form, necessitating the application of condition 2. Hence, the specific conditions under which condition 2 applies are heavily dependent on the astrophysical environment, such as the baryon number density and the expansion rate of the universe. When either condition 1 or 2 is applicable, $\sigma_{\rm eff} = \sigma_{\rm geo}$ must be used.

\subsubsection{Cooling time}
\label{cooling_time}

We now possess all the essential equations required to calculate the AQN temperature; however, the thermal equilibrium condition, crucial for applying Eq. (\ref{eq:dFtot}), must be validated. The thermal equilibrium can be established and maintained only if the AQN cooling time, $t_{\rm cool}$, exceeds the mean collision time, $t_{\rm coll} \sim \Gamma^{-1}$, as specified in Eq. (\ref{eq:Gamma}). Using the previous equations and assumptions, we now demonstrate that the framework consistently supports $t_{\rm cool} \gg t_{\rm coll}$.

The cooling time is calculated by solving Eq. (17) from \cite{Ge:2020cho}:

\begin{equation}
\frac{\rmd T_{\rm AQN}(t)}{\rmd t} = -\frac{3 F_{\rm tot}^{\rm th}(T_{\rm AQN})}{R_{\rm AQN} \, c_{\rm v}(T_{\rm AQN})}.
\label{eq:cooling}
\end{equation}
Here, $c_{\rm v}$ represents the specific heat of the core material, defined as:

\begin{equation}
c_{\rm v} \simeq \frac{1}{3} T_{\rm AQN} (\mu_d^2 + \mu_u^2).
\end{equation}

In this formula, the chemical potentials $\mu_d$ and $\mu_u$, each approximately 500 MeV, correspond to the nuclear material in the color superconducting phase. Solving Eq. (\ref{eq:cooling}) determines the cooling time scale, $t_{\rm cool}$:

\begin{equation}
t_{\rm cool} = \frac{\pi \mu_{u,d}^2 R_{\rm AQN} m_e^{1/4}}{54 \alpha^{5/2} T_0^{9/4}}.
\label{eq:t_cool}
\end{equation}

Here, $T_0$ represents an arbitrary initial temperature of the AQN at $t=0$. Within a cosmological context, the typical range of AQN temperatures is $T_{\rm AQN} \in [10^{-2}, 1]{\rm eV}$. Calculations of $t_{\rm cool}$ with $T_0$ in this range yield cooling times between $10^{12}\,{\rm s}$ and $10^8\,{\rm s}$, which is approximately four orders of magnitude larger than the average time between two collisions, $\Gamma^{-1}$. Hence, we conclude that the AQN positron sphere consistently remains in thermal equilibrium, verifying that Eq. (\ref{eq:dFtot}) is an excellent approximation for the considered range of $n_{\rm b}$ and $T_{\rm gas}$.

\subsubsection{Calculation of the AQN temperature $T_{\rm AQN}$}
\label{Taqn_calc}

The electrosphere temperature, $T_{\rm AQN}$, can be calculated by combining Eqs. (\ref{eq:E_ann}) and (\ref{eq:equi_condition}) and using Eq. (\ref{eq:Reff}) for $\sigma_{\rm eff}$. The resulting $T_\aqn$, when neither conditions 1 nor 2 apply, is given by:

\begin{equation}
\left. T_\aqn \right|_{\rm eff} = m_e \left[\frac{2{\rm GeV}~f(1-g)}{8\alpha^{3/2}T_{\rm gas}^2}\,3\pi n_{\rm b}\,\Delta{\rm v}\,R_{\rm AQN}^2\right]^\frac{4}{7}\,.
\label{eq:Taqn_with_Reff}
\end{equation}
Note that it is typically appropriate to set $f=1$ (since all captured charged particles will eventually annihilate), but we retain $f$ in the formula for clarity. Additionally, we use the subscript \texttt{eff} to signify that this expression is valid only when Eq. \eqref{eq:Reff} holds.
This equation can also be expressed as a product of dimensionless ratios, as detailed in Table \ref{tab:Eqs_cgs2nat} in Appendix \ref{app: units}:

\begin{eqnarray}
    \left. T_{\rm AQN}\right|_{\rm eff} &=& 1.4\times 10^3~{\rm eV} \left(\frac{n_{\rm b}}{1{\rm\, cm^{-3}}}\right)^\frac{4}{7} \left(\frac{\Delta{\rm v}}{10^{-3}c}\right)^\frac{4}{7} \left(\frac{f(1-g)}{0.9}\right)^\frac{4}{7} \times \nonumber \\
    &&\left(\frac{R_\aqn}{2.25\times 10^{-5}{\rm cm^{-3}}}\right)^\frac{8}{7} \left(\frac{{1\rm\,eV}}{T_{\rm gas}}\right)^\frac{8}{7}\,.
    \label{TAQN_eff}
\end{eqnarray}
When conditions 1 or 2 apply, Eq. \eqref{eq:Reff} does not hold anymore, and we must use $\sigma_{\rm eff} = \sigma_{\rm geo}$, the AQN temperature is then given by:

\begin{equation}
\left. T_\aqn\right|_{\rm geo} = \left[\frac{3\pi}{4} \frac{2~{\rm GeV}}{16\alpha^{5/2}} f(1-g) m_e^{1/4} \Delta{\rm v}~n_{\rm b}\right]^{4/17}\,.
\label{eq:Taqn_with_Rgeo}
\end{equation}
The subscript \texttt{geo} indicates that Eq. (\ref{eq:Taqn_with_Rgeo}) specifically pertains to the geometrical cross-section.

Note that it is possible to combine Eqs. \eqref{eq:Taqn_with_Reff} and \eqref{eq:Taqn_with_Rgeo} into a single expression for $T_\aqn$ which works for both the effective and geometrical cases, if the right hand side is expressed as a function of $\sigma_{\rm eff}/\sigma_{\rm geo}$:

\begin{equation}
T_\aqn = \left[\frac{3\pi}{4}\frac{2~{\rm GeV}}{16\alpha^{5/2}}f(1-g)m_e^{1/4}\Delta{\rm v}~n_{\rm b}\left(\frac{\sigma_{\rm eff}}{\sigma_{\rm geo}}\right)\right]^{4/17}\,.
\label{eq:Taqn_with_Reff_alt}
\end{equation}
This equation is useful for directly calculating $T_\aqn$ from Eq. \eqref{eq:Reff}, as long as we use $\sigma_{\rm eff}=\sigma_{\rm geo}$ if either condition 1 or 2 applies. One can see that Eq. \eqref{eq:Taqn_with_Reff_alt} is identical to Eq. \eqref{eq:Taqn_with_Rgeo} when $\sigma_{\rm eff}=\sigma_{\rm geo}$, and identical to Eq. \eqref{eq:Taqn_with_Reff} otherwise.

\begin{figure}
    \begin{center}
	\includegraphics[width=13cm]{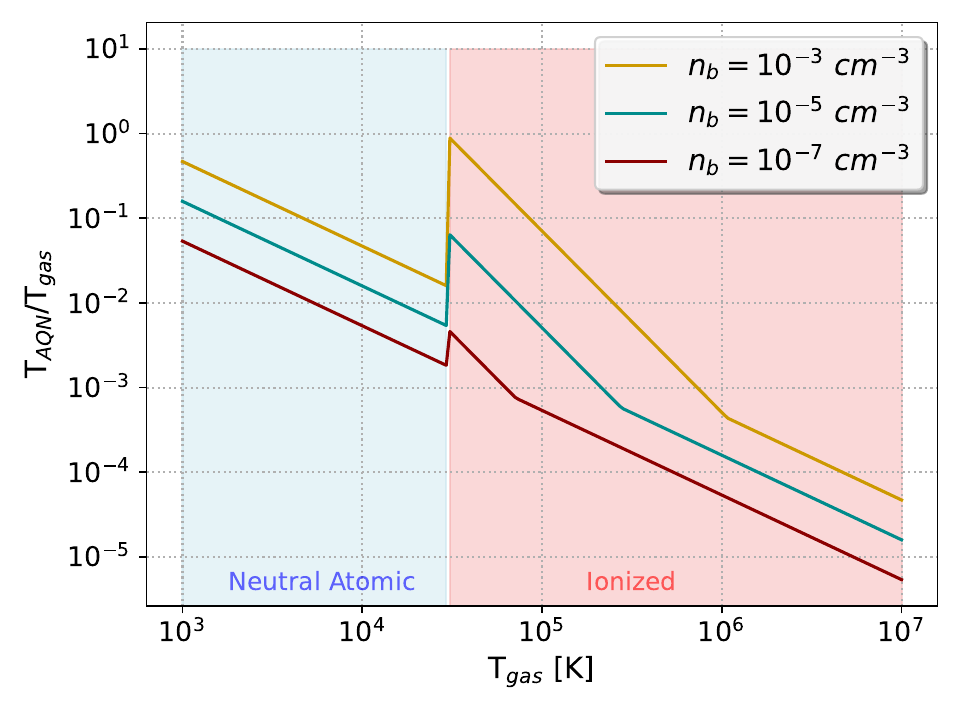}
	\end{center}
    \caption{AQN temperature over baryon gas temperature, $T_{\rm AQN} / T_{\rm gas}$, as a function of $T_{\rm gas}$  and for different baryon number density $n_{\rm b}=[10^{-3},10^{-5},10^{-7}]~\rm cm^{-3}$ for the yellow, blue and red lines respectively. The nugget mass is $m_{\rm AQN}=10{\rm\,g}$ and $\Delta {\rm v}=10^{-3}c$. The jump in all curves at $T_{\rm gas}=3\times10^4{\rm\,K}$ corresponds to the rapid decrease of the effective cross-section $\sigma_{\rm eff}$ down to $\sigma_{\rm geo}$ when the baryonic environment is no longer ionized. The choice of $3\times10^4{\rm\,K}$ was chosen arbitrarily and only serves for illustration purposes, its real value depends on the specific environment. Each curve also shows a high $T_{\rm gas}$ transition, which comes from the fact that the effective cross-section can never decrease below the geometrical cross-section.}
    \label{fig:TAQN_Tgas_ratio}
\end{figure}

Figure \ref{fig:TAQN_Tgas_ratio} displays the ratio $T_{\rm AQN} / T_{\rm gas}$ using Eq. \eqref{eq:Taqn_with_Reff_alt} as a function of $T_{\rm gas}$ for three different values of $n_{\rm b}$. On this Figure, we arbitrarily choose a sharp transition from ionized to neutral atomic baryonic gas at $T_{\rm gas} = 3 \times 10^4 {\rm K}$ for illustration purposes.
Each curve is divided into three segments with varying slopes: the leftmost segment, under $T_{\rm gas} =3 \times 10^4 {\rm K}$, where condition 2 would apply, indicating a neutral atomic environment; a middle segment where Eq. \eqref{eq:Reff} is applicable; and a rightmost segment at high $T_{\rm gas}$ where condition 1 applies. The segments under conditions 1 and 2 exhibit identical slopes.

\subsubsection{The Electrosphere Spectral Emissivity}
\label{sec:Fnu_th}

The emission spectrum $F_\nu^{\rm th}(T_{\rm AQN})$ of the electrosphere was derived in \cite{Forbes:2008uf}:

\begin{equation}
F_\nu^{\rm th}(T_{\rm AQN}) = \frac{8}{45} T_{\rm AQN}^3 \alpha^{5/2} \left(\frac{T_{\rm AQN}}{m_e}\right)^{1/4} H\left(\frac{2\pi\nu}{T_{\rm AQN}}\right)\,.
\label{eq:dFnu}
\end{equation}
Here, the function $H(x)$ is defined as:

\begin{equation}
H(x)\simeq \begin{cases}
(1+x)e^{-x}(17-12\ln(x/2))\,, & \text{if } x < 1\,;\\
(1+x)e^{-x}(17+12\ln(2))\,, & \text{if } x \ge 1\,.
\end{cases}
\end{equation}
where the term $2\pi\nu$ in $H(x)$ denotes the angular frequency, expressed in natural units (see Table \ref{tab:unit conversion} in Appendix \ref{app: units})\footnote{In natural units, it is common to use the angular frequency $\omega$, such that in C.G.S. systems, $h\nu$ is replaced by $\hbar\omega$, consistent with the use of $\hbar = 1$. However, in astrophysics, the frequency $\nu$ is typically used because it can be directly compared to the observational capabilities of radio telescopes.}. The exponential dependence on frequency $\nu$ in Eq. \eqref{eq:dFnu} is characteristic of the thermal Bremsstrahlung emission. However, the low-frequency behavior deviates slightly from conventional Bremsstrahlung, influenced by the geometry and size of the positron cloud, as noted in \citep{Forbes:2008uf}. Below $\nu \sim 1~\rm GHz$, Eq. \eqref{eq:dFnu} becomes invalid and more theoretical work is necessary to extend it to lower frequencies. For the purposes of this work, we will disregard this limitation as it occurs at frequencies outside our range of interest.

The C.G.S. unit for $F_\nu^{\rm th}(T_{\rm AQN})$ is $\rm erg~s^{-1}~cm^{-2}~Hz^{-1}$. For consistency, we should note that Eq. \eqref{eq:dFtot} is the integral of Eq. \eqref{eq:dFnu} over $\nu$:

\begin{equation}
\label{eq:dF_tot(T)}
F_{\rm tot}^{\rm th}(T_{\rm AQN}) = \int_0^\infty \rmd\nu\, F_\nu^{\rm th}(T_{\rm AQN})\,.
\end{equation}

\begin{figure}
    \begin{center}
	\includegraphics[width=13cm]{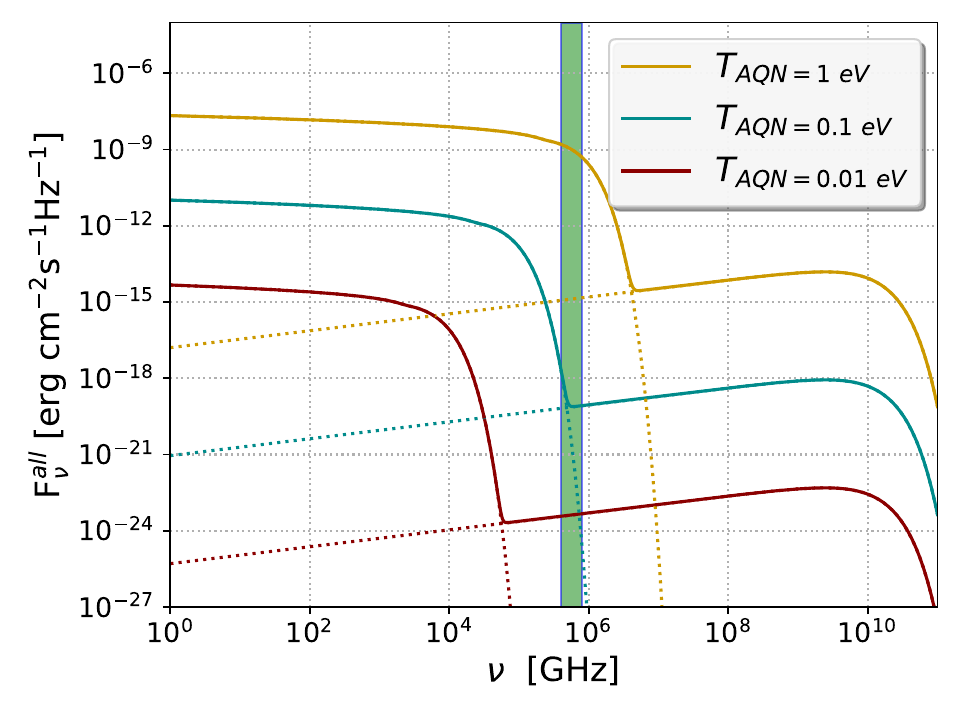}
	\end{center}
    \caption{The solid lines show the total emission spectrum $F_\nu^{\rm all} = F_\nu^{\rm th}+ F_\nu^\textrm{non-th}$ for AQNs at different temperatures $T_\aqn=[1,10^{-1},10^{-2}]~\rm eV$ for the yellow, blue and red lines respectively. Each solid line is the sum of thermal [Eq. (\ref{eq:dFnu})] and non-thermal [Eq. (\ref{eq:dFnu_xray ::2})] emissions.  The dotted lines show the separation between the thermal and non-thermal components. The thermal contribution dominates at low frequency in radio, while the non-thermal contribution dominates at high frequency in X-ray. The vertical green region represents the optical range of frequencies. One can see how the emission spectrum in the optical range is highly dependent on the AQN temperature.}
    \label{fig:dFnu_tot}
\end{figure}

Figure \ref{fig:dFnu_tot} displays $F_\nu^{\rm th}(T_{\rm AQN})$ for different values of $T_\aqn$, along with the non-thermal emission, which will be calculated in Section \ref{Fnu_nonth}. In the radio range, the amplitude of the spectrum scales approximately as $\sim T_\aqn^{3.25}$. However, in the infrared-UV range, the amplitude is highly dependent on $T_\aqn$ due to the exponential factor. This characteristic is particularly interesting for designing tests of the model, which will be further discussed in the conclusion section.

In order to evaluate the radiation emanating from a spatial region containing a number density $n_{\rm AQN}$ of AQNs, one must compute the so-called spectral emissivity $\epsilon_\nu^{\rm th}(T_{\rm AQN})$ emitted from AQN-baryon collisions per unit time, volume, and frequency, which is given by:

\begin{equation}
\epsilon_\nu^{\rm th}(T_{\rm AQN}) = 4\pi R_{\rm AQN}^2~ F_\nu^{\rm th}(T_{\rm AQN})~n_{\rm AQN}\,.
\label{eq:epsilon_nu}
\end{equation}
The C.G.S. unit for $ \epsilon_\nu^{\rm th}(T_{\rm AQN})$ is $\rm erg\,s^{-1}\,cm^{-3}\,Hz^{-1}$.

In summary, we have outlined the methodology to calculate the thermal emissivity spectrum of the AQN dark matter, once the macroscopic parameters—specifically the AQN number density $n_\aqn$, the baryon number density $n_{\rm b}$, the gas temperature $T_{\rm gas}$ and the relative velocity between dark matter and baryons—are known. The only fundamental free parameter\footnote{There are many parameters which enter our computations such as $f$ or $g$. However, these parameters are not fundamental parameters of the system, and in principle could be computed from first principles. Such computations, however, are not feasible at this moment, as it requires good understanding of the strongly coupled QCD.} of the model is the AQN mass $m_\aqn$ since it determines its size $R_\aqn$. It is expected that the AQN mass should follow a power law distribution, but in this work we consider a fixed value of the mass for simplicity as explained  in Sect.\ref{sect:mass-distribution}.

\subsection{The Non-Thermal Emission}
\label{Fnu_nonth}

As discussed in Section \ref{sec:energy_budget}, the collision of an antimatter AQN with a baryon also results in non-thermal emission, manifesting as a short pulse of predominantly X-ray photons in the $\sim$ 1-10 keV range \citep{Forbes:2006ba}. This non-thermal emission originates from the Bremsstrahlung of relativistic positrons at the impact site by the mean field of surrounding positrons. Unlike the thermal process, this emission does not first involve heat transfer through the anti-quarks core, which allows the radiation to escape immediately and results in the short duration of the pulse. For a collision with a proton, the total amount of energy per event is given by:

\begin{equation}
\Delta E_{\rm ann}^\textrm{non-th}
\approx 2~{\rm GeV}fg
\end{equation}
%
%Similarly, the energy emitted in the form of x-rays per unit of time, analogous to Eq. \eqref{eq:E_ann}, is expressed as Eq. (\ref{eq:E_Xray}):
%
%\begin{equation}
%\label{E_ann x-ray}
%\frac{{\rm d} E_{\rm ann}^{\rm non-th}}{{\rm d} t}
%\approx 2~{\rm GeV}fg\sigma_{\rm eff}\Delta \rm v~n_{\rm b}
%\end{equation}
%
The chemical potential of these positrons, located deep within the electrosphere, is approximately $\mu\sim10\rm{\,MeV}$. They emit X-rays through Bremsstrahlung, characterized by a qualitative spectrum:

\begin{equation}
\label{dFnu x-ray}
F_\nu^\textrm{non-th}
\propto\frac{\nu}{\nu_{\rm c}}
\int_{\nu/\nu_{\rm c}}^\infty K_{5/3}(x)\, \rmd x\,,
\end{equation}
where $K_{5/3}(x)$ is the modified Bessel function of the second kind, and $x=\nu/\nu_c$. The characteristic angular frequency $2\pi\nu_{\rm c}$ is approximately 30 keV, as derived from estimates of the scattering timescale \citep{Forbes:2006ba}.
The derivation of Eq. \eqref{dFnu x-ray} is based on classical electrodynamics assuming an instantaneous  average radius of curvature of the positrons trajectories described by a gaussian distribution, which corresponds to the characteristic frequency $\nu_c$. In reality, the positrons energy distribution is more complicated and a full quantum mechanical approach should also be considered. This is left for a future work, and in this paper we will assume that a single characteristic frequency $\nu_c$ is a valid approximation for our purpose. Nevertheless, it is important to keep in mind that the non-thermal spectrum calculation is less robust than the thermal emission spectrum outlined in Eq. \eqref{eq:dFnu}.

Integrating over the modified Bessel function as specified in Eq. \eqref{dFnu x-ray} provides a simple analytical approximation:\footnote{It should be noted that the general form of the non-thermal spectrum can be expressed as:
\begin{equation*}
F_\nu^\textrm{non-th}
\propto x^\beta e^{-x}\,
\end{equation*}
where the Bremsstrahlung emission is dependent on the energy distribution of the positrons. Here, we adhere to the convention $\beta=1/3$, as utilized in the original paper \cite{Forbes:2006ba}, which assumes that the positron beams are highly correlated.}

\begin{equation}
x\int_x^\infty K_{5/3}(x')\,\rmd x'
\approx 1.81x^{1/3}e^{-x}\,.
\end{equation}

This leads us to define a formula analogous to Eq. \eqref{eq:dF_tot(T)}:
\begin{equation}
F_{\rm tot}^\textrm{non-th}
=\int_0^\infty \rmd\nu\,F_\nu^\textrm{non-th}\,.
\label{eq:F_tot_nonth}
\end{equation}
An expression for $F_\nu^\textrm{non-th}$ can be obtained with the following argument: the non-thermal radiation is emitted in a short pulse for a single collision event, but given the large number of collisions per unit of time within cosmologically large volumes, we can use Eq. (\ref{eq:E_Xray}) as an estimate of the average power emitted non-thermally.
%
%\begin{equation}
%\label{E_ann x-ray}
%\frac{d E_{\rm ann}^{\rm non-th}}{d t}
%\approx 2{\rm\,GeV}fg\sigma_{\rm eff}\,\Delta {\rm v}\,n_{\rm b}\,.
%\end{equation}
%
The average power, per AQN, is then simply given by:
\begin{equation}
\label{E_ann x-ray}
\frac{\rmd E_{\rm ann}^\textrm{non-th}}{\rmd  t}
= 4\pi R_\aqn^2  F_{\rm tot}^\textrm{non-th}\,,
\end{equation}
where
\begin{subequations}
\begin{equation}
F_\nu^\textrm{non-th}
\approx \frac{2 {\rm\,GeV}fg}{4\,\Gamma({4/3})}\frac{n_{\rm b}\Delta {\rm v}}{\nu_c}
\left(\frac{R_{\rm eff}}{R_\aqn}\right)^2
\left(\frac{\nu}{\nu_c}\right)^{\frac{1}{3}}e^{-\nu/\nu_c},
\label{eq:dFnu_Xray}
\end{equation}
and the spectral emissivity is:

\begin{equation}
\epsilon_\nu^\textrm{non-th}
= 4\pi R^2 F_\nu^\textrm{non-th} n_{\rm AQN},
\end{equation}
\end{subequations}
where the Gamma function $\Gamma(4/3) \sim 0.893$. Additionally, Eq. (\ref{eq:dFnu_Xray}) can be recast as a function of $T_{\rm AQN}$ using Eq. \eqref{eq:Taqn_with_Reff_alt}:

\begin{equation}
\label{eq:dFnu_xray ::2}
F_\nu^\textrm{non-th} = \frac{16 \alpha^{5/2}}{3 \pi\, \Gamma(4/3)}\frac{g}{1-g}\frac{1}{\nu_c}\frac{T_{\rm AQN}^{17/4}}{m_e^{1/4}} \left(\frac{\nu}{\nu_c}\right)^{\frac{1}{3}}e^{-\nu/\nu_c}\,.
\end{equation}
Figure \ref{fig:dFnu_tot} shows the non-thermal X-ray spectral emissivity for various AQN temperatures, along with the thermal emission calculated in Section \ref{sec:Fnu_th}.

Note that, in addition to X-rays, other non-thermal electromagnetic radiations such as the direct production of 0.1-1 GeV gamma rays and a 511 keV line from electron-positron annihilation are possible, though less prominent \citep{Forbes:2006ba,Oaknin:2004mn,Zhitnitsky:2006tu,Lawson:2007kp}. Our focus in this work is primarily on the dominant thermal and non-thermal emissions, as derived above.

\section{Simulations}
\label{sec:sims}

In Section \ref{sec:emission_mec} we described the calculation of the spectral emissivity $\epsilon_\nu^{\rm th}$ and $\epsilon_\nu^\textrm{non-th}$ for the thermal and non-thermal processes, given the AQN number density $n_\aqn$, baryon number density $n_{\rm b}$ and temperature $T_{\rm gas}$ and the average speed between dark matter and baryons $\Delta\rm v$. The AQN specific intensity $I_\nu(\vec\theta )$ at a given position $\vec\theta$ on the sky is given by the sum of the spectral emissivity along redshift for each line-of-sight in direction $\vec\theta$. Hydrodynamical simulations provide all the physical quantities needed as a function of 3D position in space $\vec r$, for different redshifts $z$, from which the sky projection can be performed. 

The procedure is as follows: from the dark matter mass density, $\rho_{\rm DM}(\vec r,z)$, we calculate the number density of AQNs, $n_\aqn(\vec r,z)$. This is then combined with the baryons' number density, $n_{\rm b}$, the gas temperature, $T_{\rm gas}$, and the relative speed, $\Delta\rm v$, to determine the AQN temperature, $T_\aqn(\vec r,z)$. From there, the spectral emissivity, $\epsilon_\nu(\vec r,z)$, is calculated. A criterion to fix the ionization level of the baryonic matter also needs to be established. This section outlines the computation of $\epsilon_\nu(\vec r,z)$ and details the construction of the light-cone and the calculation of the specific intensity $I_\nu(\vec\theta)$.

\subsection{Data cubes}
\label{subsec:datacubes}

\begin{figure}
\centering
\noindent\makebox[\textwidth]{%
  \hspace{-10pt}%
  \includegraphics[width=1.1\textwidth]{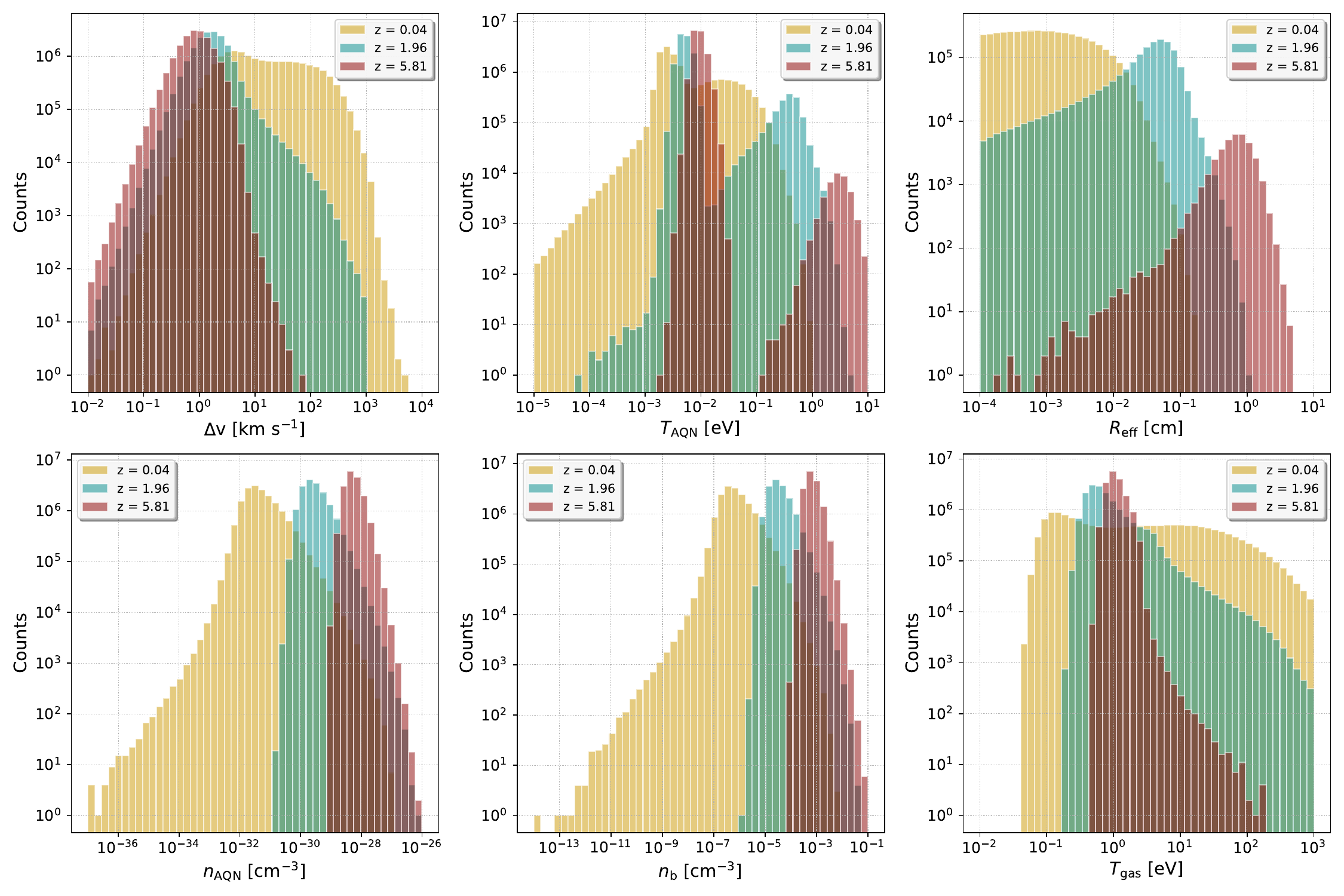}
}
\caption{Histograms of the physical quantities relevant for the AQN emission for three different redshifts $z=[0.04,1.96,5.81]$ {\bf measured in the simulations}. {\bf The histogram plots show the number of voxels in the simulation that fall into each bin.}. This is for an AQN mass of $m_\aqn=100~\rm g$, which correspond to an AQN radius of $R_{\rm geo}=4.7\times 10^{-5}~{\rm cm}$. Other AQN masses are not display in order to avoid a confusing plot. We have assumed a mixed ionized case with an ionization temperature $T_{\rm gas}=2.6~\rm eV$ ($3\times 10^4~\rm K$). From the top-left to the bottom-right panels: $\Delta{\rm v}$ is the relative speed between baryons and dark matter, $T_\aqn$ is the AQN electrosphere temperature, $R_{\rm eff}$ is the tail of the effective radius distribution (for most AQNs we have $R_{\rm eff}=R_{\rm geo}$), $n_\aqn$ is the number density of AQNs, $n_b$ is the number density of baryons (assuming protons) and $T_{\rm gas}$ is the gas temperature.}
\label{fig:cube_properties}
\end{figure}

We utilized the {\it Box3/hr} from the \textit{Magneticum Pathfinder} suite of cosmological hydrodynamical simulations \citep{2016MNRAS.463.1797D}, which covers a comoving cosmological volume of $(128 h^{-1}c{\rm\,Mpc})^3$ with $576^3$ dark matter particles and an equal number of gas particles and periodic boundary conditions. The \textit{Magneticum} simulations adopt the WMAP cosmology \cite{2009ApJS..180..330K} with a Hubble constant of $H_0 = 70.4$ ${\rm\,km\,s^{-1}\,Mpc^{-1}}$, total matter density parameter $\Omega=0.272$, a baryonic fraction of 16.8\%, index of the primordial power spectrum $n = 0.963$, and normalization of the fluctuation amplitude $\sigma_8=0.809$. More information about the \textit{Magneticum} project is available at \url{www.magneticum.org}.

The \textit{Magneticum} simulations were conducted using the Tree/SPH code \textsc{P-Gadget-3}, an advancement of the publicly available \textsc{P-Gadget-2} \citep{2005MNRAS.364.1105S}. It incorporates an enhanced Smoothed Particle Hydrodynamics (SPH) solver \citep{2016MNRAS.455.2110B} that includes improvements in numerical viscosity \citep{2005MNRAS.364..753D}. The simulations account for a range of physical processes such as metallicity-dependent gas cooling \citep{2009MNRAS.393...99W}, star formation and stellar feedback \citep{2005MNRAS.361..776S, 2003MNRAS.339..289S}, UV background \citep{2001cghr.confE..64H}, chemical enrichment \citep{2007MNRAS.382.1050T, 2004MNRAS.349L..19T}, and thermal conduction at 1/20th of the Spitzer level \citep{2014arXiv1412.6533A}. Additionally, the simulations monitor the evolution of supermassive black holes (SMBHs) and their associated feedback, building upon the foundational model by \cite{2005MNRAS.361..776S} with subsequent enhancements by \cite{2010MNRAS.401.1670F} and \cite{2014MNRAS.442.2304H}. The data cubes are $256^3$ voxels and are given for eight time stamps which correspond to redshifts $z_i=[0, 0.252, 0.471, 1.18, 1.98, 2.792, 4.23, 5.34]$.

Since the key quantity for calculating the specific intensity $I_\nu(\vec \theta)$ is the spectral emissivity $\epsilon_\nu(\vec r,z_i)$, we first compute $\epsilon_\nu$ for the initial eight redshifts, and then construct the light-cone. The quantities directly obtained from the simulations are $\Delta{\rm v}$, $n_{\rm b}$, $T_{\rm gas}$, and $n_\aqn$ (assuming a specific AQN mass for the latter). Conversely, the  quantities derived from the simulations, $T_\aqn$ and $R_{\rm eff}$, require knowledge of the baryon gas's ionization level, since the AQN signal is strongly dependent on the ionization level. The highest redshift of the simulation being $z=5.34$, we know the Universe should be fully reionized, therefore the gas should be ionized in the entire light cone. In reality, the exact history of reionization and how it happens is still unknown, and moreover, as the Universe continues expanding and the galaxies and stars to form, there are regions where the gas cools down and returns to atomic form. How this is happening is still an active field of research. Therefore, as a conservative approach, we will consider three ionization levels in the following: 100\% ionized, a mixed case where we assume that the gas below $T_{\rm gas}=2.6{\rm\,eV}$ is in atomic form, and a fully neutral atomic case. Since it is expected that the Universe is in a highly ionized state, the fully neutral atomic case is not realistic, but we use it as a benchmark representing what the lowest AQN signal would be if the cross-section was only geometrical $\sigma_{\rm geo}$.

Figure \ref{fig:cube_properties} shows the physical quantities relevant for the AQN emission for three different redshifts and for the mixed ionization case. The most interesting feature is the double structure histogram for $T_\aqn$. This is coming from the gas temperature cut $T_{\rm gas}=2.6{\rm\,eV}$ below which the baryons surrounding the AQNs are assumed in neutral atomic form. For the same baryon number density $n_b$ and relative speed $\Delta{\rm v}$, AQNs in a neutral atomic environment are cooler than the AQNs in a fully ionized environment by approximately two orders of magnitude. This separation is particularly strong at high redshift, but at lower redshift the double structure eventually merges in a single histogram with a bump in the middle, and at redshift $z=0.04$, the average and dispersion of the $T_\aqn$ is mostly dominated by the AQN in the ionized environment. 
The top-right panel of Figure \ref{fig:cube_properties} shows that in the mix ionized case, the majority of the AQNs have $R_{\rm eff}=R_{\rm geo}$ (the panel only shows the tail of the $R_{\rm eff}$ distribution, see the caption). We can use Figure \ref{fig:cube_properties} to anticipate what should happen in the fully ionized and fully neutral atomic cases (not shown): for the fully ionized case, only the high $T_\aqn$ histogram is populated, while in the fully neutral atomic case, only the low $T_\aqn$ histogram is populated. The impact on the observability of the AQN signal will be discussed later.
In an ionized environment, all the AQNs have an effective radius $R_{\rm eff} \gg R_{\rm geo}$ at the highest redshift of our light-cone, but at lower redshift the distinction between the fully ionized and the mix cases is not strong.

\begin{figure}
\centering
\includegraphics[width=1.\textwidth]{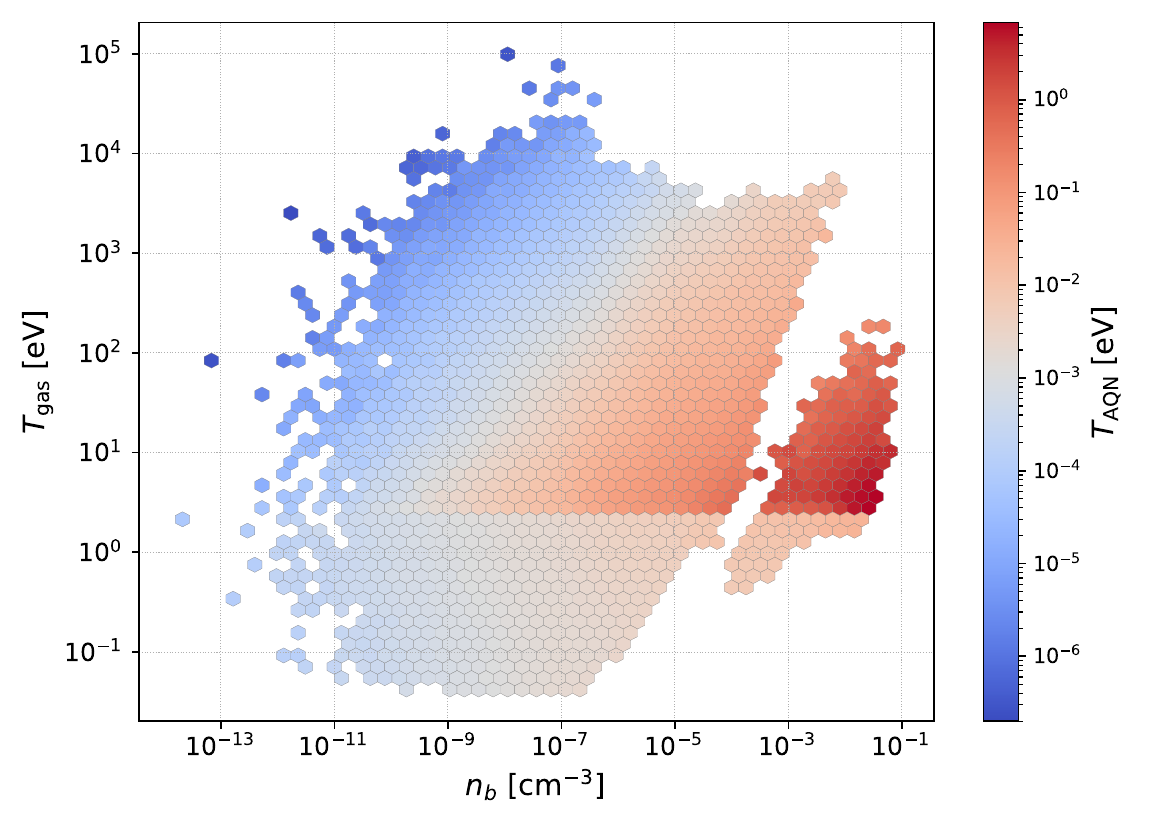}
\caption{Average AQN temperature $T_\aqn$ {\bf (in eV)} as a function of the baryon number density $n_b$ and the gas temperature $T_{\rm gas}$ for two different redshifts, {\bf $z=0.04$ (corresponding to the larger group of points at the center) and  $z=5.81$ (corresponding to the smaller group of points on the right). We use a single color map for the two redshifts.} The AQN physical properties are the same as in Figure \ref{fig:cube_properties}.}
\label{fig:Taqn_scatter_plot}
\end{figure}

Figure \ref{fig:Taqn_scatter_plot} shows clearly the distinction between the neutral atomic environment ($T_{\rm gas}< 2.6~\rm eV$) and ionized environment ($T_{\rm gas}> 2.6~\rm eV$), and the sharp jump in $T_\aqn$ is what can also be seen in Figure \ref{fig:TAQN_Tgas_ratio} for various baryon number densities.

\subsection{Calculation of the specific intensity $I_\nu$}
\label{subsec:Computing I_nu}

\begin{figure}
\centering
\includegraphics[width=0.85\textwidth]{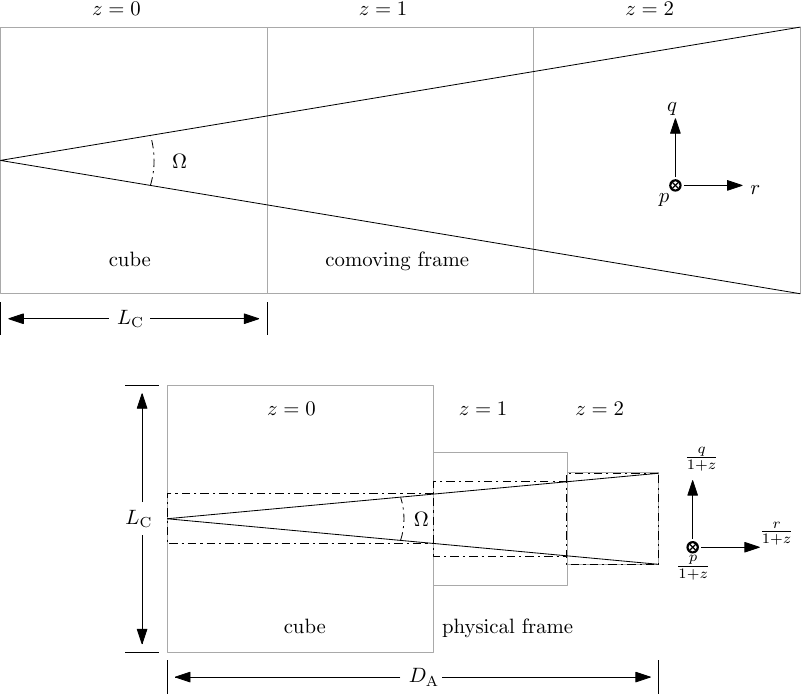}
\caption{Illustrative construction of cubes in comoving (upper) and physical (lower) coordinates. $(p,q,r)$ are the Cartesian comoving coordinates with $r$ extending along the redshift axis. $D_{\rm A}=(1+z)^{-2}D_{\rm L}$ is the angular diameter distance. $L_{\rm C}=128h^{-1}\rm Mpc$ is the comoving size of a cube. In a continuous space, the observer has the same solid angle $\Omega$ both in the comoving and physical frames. In the simulation, the space is divided into a series of subsequent cubes at discrete redshifts. The projection within each cube along $\Omega$ becomes parallel (dash-dotted), instead of conical (solid).}
\label{fig:cubes_merged}
\end{figure}

The calculation of $I_\nu(\vec\theta)$ first requires the spectral emissivity $\epsilon_\nu(\vec r,z_i)$ within each cube, from which we subsequently construct the light cone. All the calculations are performed for three different average AQN masses $m_{\rm AQN}=(10{\rm\,g}, 100{\rm\,g}, 1000{\rm\,g})$, which correspond to an average radius $R_{\rm AQN}=(1.9\times 10^{-5}{\rm cm}, 4.1\times 10^{-5}{\rm cm}, 8.8\times10^{-5}{\rm cm})$, and for the three ionization levels (fully ionized, mix case and neutral atomic). 

It is more convenient to use comoving distances for the calculations, the corresponding quantities are labelled with the subscript C.
The spectral emissivity $\epsilon_{\nu,\rm C}(\vec r,z_i)$ at redshift $z_i$ of the AQN-gas interaction, at comoving position $\vec r$ is given by:

\begin{equation}
    \epsilon_{\nu,\rm C} (\vec r,z_i) = 4 \pi R_{\rm AQN}^2\, F_\nu(T_\aqn(\vec r,z_i))\,n_{{\rm AQN},\rm C}(\vec r,z_i) 
    \label{eq:epsilon_nu}
\end{equation}
where $n_{\aqn ,\rm C}$ is the comoving AQN number density. $F_\nu(T_\aqn(\vec r,z_i))$ is the emission spectrum, calculated using physical quantities at redshift $z_i$. $F_\nu(T_\aqn(\vec r,z_i))$ is given by Eq. \eqref{eq:dFnu} or Eq. (\ref{eq:dFnu_Xray}) for the thermal or non-thermal process, respectively, or it can be the sum of both. The unit of $\epsilon_{\nu,C}$ is $\rm erg\,s^{-1}Hz^{-1}cm^{-3}$. Note that $ F_\nu$ and $n_{\aqn ,\rm C}$ are calculated for each voxel $\vec r$ of a data cube at redshift $z_i$. The AQN temperature $T_\aqn(\vec r,z_i)$ is calculated using the procedure described in Section \ref{Taqn_calc}.

\begin{figure}
    \begin{center}
	\includegraphics[width=13cm]{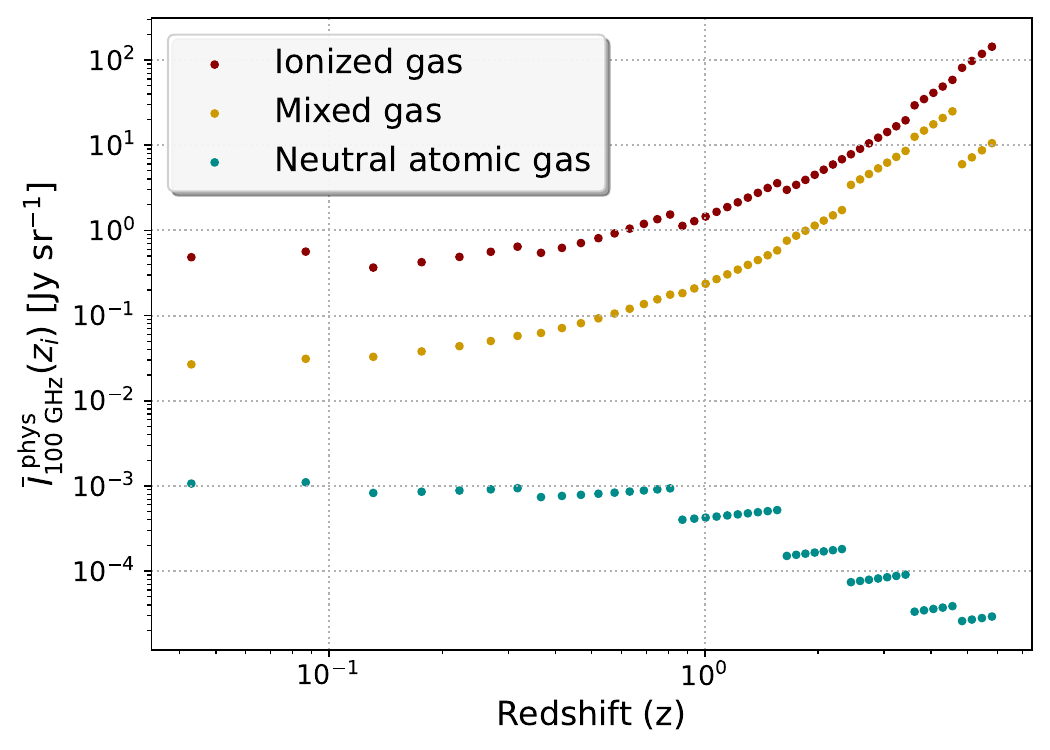}
	\end{center}
    \caption{Average intensity $\bar{I}^{\rm phys}_{\rm 100~GHz} (z_i)$ as a function of redshift $z_i$ (see Eq. \ref{eq:delta_Inu}) for an AQN mass $m_\aqn = 10~\rm g$. Each dot indicates the average intensity in the data cube at a particular redshift $z_i$. This plot shows which redshift contributes the most to the observed intensity depending on the ionization level of the environment (ionized, mixed or neutral atomic). The origin of the successive jumps in redshift is explained in the text: between two jumps, the same time stamp was used to generate the data cubes.}
    \label{fig:redshift_Inu}
\end{figure}

In order to calculate the specific intensity, $I_\nu(\vec \theta)$, we must first construct a light cone of $\epsilon_{\nu ,\rm C} (\vec r,z)$ as a continuous function of redshift and where the large scale structures do not appear at the same position in the cubes as the redshift varies. The gaps between the sampled redshifts $z_i$ are simply filled with copies of the nearest cube in comoving coordinates. For the redshift range $z=[0,5.8]$, 49 cubes are needed. The cubes are adjacent to each other, do not overlap, and constitute the basis for our light cone. In order to eliminate the spatial correlation between cubes, we employ a technique commonly used in N-body simulation, which consists in randomly shifting and rotating each cube. The 49 adjacent data cubes are constructed from the original eight time stamps. We decided to not perform any interpolation between time stamps, and instead, for any redshift $z_i$, use the nearest time stamp as the data cube. This procedure results in a quantized distribution of average cosmological values, but it will not significantly affect our results. Figure \ref{fig:cubes_merged} shows the cubes arrangements of the light cone from which the sky projection is later constructed, and a comparison between the physical and comoving frames.

Along a given line-of-sight $\vec\theta$ subtends a solid angle $\rmd  \Omega$, and the comoving volume element ${\rm dV_C}=(1+z)^3 {\rm dV}$ is given by:

\begin{equation}
{\rm d}V_{\rm C}
=(1+z)^3 \frac{{\rm d}p}{1+z}\, \frac{{\rm d}q}{1+z}\, \frac{{\rm d}r}{1+z}\approx (1+z)^3 {\rm d}\Omega\, D_{\rm A}^2(z) \frac{{\rm d}r}{1+z}
\,,
\end{equation}
where $\vec r=(p,q,r)$ are Cartesian comoving coordinates \footnote{The use of Cartesian coordinates is justified by the small angle approximation, and $(p,q,r)$ are expressed in physical units.} with $r$ along the redshift axis, and $D_{\rm A}(z)$ is the angular diameter distance at redshift $z$. The comoving volume can be rewritten as:

\begin{equation}
{\rm d}V_{\rm C}
\approx\frac{D_{\rm L}^2(z)}{(1+z)^2}\,{\rm d}\Omega\,{\rm d}r\,,
\end{equation}
where $D_{\rm L}(z)=(1+z)^2D_{\rm A}(z)$ is the luminosity distance at redshift $z$.
Using Eq. \eqref{eq:epsilon_nu}, we define the $\rmd L_{\nu'}(\vec r,z)$ the spectral luminosity [$\rm erg\,s^{-1}\,Hz^{-1}$] at rest-frame frequency $\nu'=(1+z)\nu$ emitted by a voxel (the source) at location $(\vec r,z)$ and observed within the solid angle $\rmd\Omega$:

\begin{equation}
%	\label{eq:name}
{\rm d}L_{\nu'}(\vec r,z)
= \epsilon_{\nu' ,\rm C}(\vec r,z)\,{\rm d}V_{\rm C}\,,
\end{equation}
and the rest frame luminosity of the source [$\rm erg\,s^{-1}$] in the physical frame is $\rmd L_{\nu'}(\vec r,z){\rm d}\nu'$.

The specific intensity, measured at frequency $\nu$ within $\rmd \nu$ and solid angle $\rmd \Omega$, can be expressed as:

\begin{equation}
\frac{\partial I_\nu(\vec r,z)}{\partial r}\,{\rm d}\nu\,{\rm d}\Omega\,{\rm d}r
=\frac{{\rm d}L_{\nu'}(\vec r,z)\,{\rm d}\nu'}{4\pi D_{\rm L}^2(z)}=\frac{\epsilon_{\nu' ,\rm C}(\vec r,z)}{4\pi } \frac{{\rm d}\nu'{\rm d}\Omega}{(1+z)^2}\,{\rm d}r.
\end{equation}

We conclude that $I_\nu$ [$\rm erg\,s^{-1}\,cm^{-2}\,Hz^{-1}\,sr^{-1}$] is given by:
\begin{equation}
\label{eq:I_nu}
\begin{aligned}
I_\nu(p,q)
%&=\int\frac{{\rm d} V_{\rm C}}{{\rm d}\Omega}\,n_{\rm AQN,C}(p,q,r)\,
%\frac{4\pi R^2F_{\nu'}}{4\pi D_{\rm L}^2(z)}\frac{{\rm d}\nu'}{{\rm d}\nu}  \\
&=\int \frac{\partial I_\nu(\vec r,z)}{\partial r}\,{\rm d}r\\
&=R_\aqn^2\int{\rm d} r \frac{1}{(1+z)^2}\,n_{\rm AQN,\rm C}(\vec r,z)\,
F_{\nu'}(\vec r,z)\cdot(1+z)  \\
&=R_\aqn^2\int\frac{{\rm d} r}{1+z}\,
n_{\rm AQN,\rm C}(\vec r,z)\,F_{\nu'}(\vec r,z)\,,
\end{aligned}
\end{equation}
Note that $r$ is implicitly a function of $z$, but we keep $z$ as an explicit variable as a marker of which data cube time stamp should be used (this is important when rewriting the integral in the discretized form). For analytical calculations, one may express $r$ as a function of $z$, but for this work we are concerned with discretized 3-dimensional quantities.
As shown in Figure \ref{fig:cubes_merged}, we have a stack of cubes with the same comoving size. Each cube contains discrete pixels of numerical values in Cartesian coordinates $\vec r_i=(p_i,q_i,r_i)$ at discrete redshift $z_i$. Then, for each cube at redshift $z_i$, the corresponding $I_\nu(p_i,q_i,z_i)$ is:

\begin{equation}
\label{eq:I_nu(z,m,n)}
I_\nu(z_i,p_i,q_i)
=\frac{R^2\delta r}{1+z_i}
\sum_{r_i} n_{\rm AQN,\rm C}(p_i,q_i,r_i)
F_{\nu'}(z_i,p_i,q_i,r_i)\,,
\end{equation}
where $\delta r$ is the comoving lattice spacing between two nearby pixels in the $r$-direction. As shown in figure \ref{fig:cubes_merged}, projections of $I_\nu(p_i,q_i,z_i)$ in each cube are parallel, instead of conical, for simplicity in simulations. For each plate $I_\nu(p_i,q_i,z_i)$, only one square slice of size $\Delta p\times\Delta q$ contributes to the overall projection of intensity background. The $\Delta p$ and $\Delta q$ are redshift-dependent with the following relation:

\begin{equation}
\begin{aligned}
\Delta p(z_i)
=\Delta q(z_i)
=(1+z_i)D_{\rm A}(z_i)\cdot\theta\,,\\
\theta
=\frac{L_{\rm C}}{(1+z_{\rm max})\,D_{\rm A}(z_{\rm max})}\,,
\end{aligned}
\end{equation}
where $D_{\rm A}(z_i)=(1+z_i)^{-2}D_{\rm L}(z_i)$ is the angular diameter distance, $\theta=\sqrt{\Omega}$ is the angle of the conical projection (see figure \ref{fig:cubes_merged}), $L_{\rm C}$ is the comoving size of a cube, $z_{\rm max}$ is the maximal redshift in the cubes. The range of $(p_i,q_i)$ for a cube at redshift $z_i$ is chosen to be:\footnote{Here we implicitly choose the projection plate $(p_i,q_i)$ locating at one corner of the $p-q$ face of the cube, instead of locating near the center of the $p-q$ face. The choice of location is unimportant because the pixels will be randomized by shifting and rotating in each cube, as discussed earlier in this subsection. }

\begin{equation}
p_i\in[0,\Delta p(z_i)]\,,\qquad
q_i\in[0,\Delta q(z_i)]\,.
\end{equation}
Beyond this range, the data points are out of the projection light cone $\Omega$ and have no impact on the intensity function $I_\nu$. 
The intensity spectrum $I_\nu(p_i,q_i,z_i)$ in the comoving frame can be converted to the physical frame:

\begin{equation}
I^{\rm phys}_\nu(z_i,x_i,y_i)
\equiv I_\nu(z_i,\frac{\Delta p(z_i)}{\Delta p(z_{\rm min})}x_i,
\frac{\Delta q(z_i)}{\Delta q(z_{\rm min})}y_i)\,,
\end{equation}
where $(x,y)$ is defined to be the physical Cartesian coordinate measured at redshift $z=z_{\rm min}$ (the minimal nonzero $z_i$ in the cubes):

\begin{equation}
(x_i,y_i)
=\left.\frac{1}{1+z}(p_i,q_i)\right|_{z=z_{\rm min}}
\end{equation}
such that the observation angle $\vec\theta$ can be specified by $(x_i,y_i)$. The total background intensity $I^{\rm phys}_{{\rm tot},\nu}(x_i,y_i)$ is obtained by stacking up all $I^{\rm phys}_\nu(z_i,x_i,y_i)$ plates:

\begin{equation}
I^{\rm phys}_{{\rm tot},\nu}(x_i,y_i)
=\sum_{z_i} I^{\rm phys}_\nu(z_i,x_i,y_i)\,.
\label{eq:delta_Inu}
\end{equation}

Figure \ref{fig:redshift_Inu} shows the scaling of the AQN signal $\bar I^{~\rm phys}_{\rm 100~GHz}(z_i)$ with redshift $z_i$ for the three ionization levels of the environment, at $\nu=100~\rm GHz$ and $m_\aqn = 10~\rm g$. At high redshift, the matter (baryons and dark matter) number density is high, but the baryon temperature is generally not (see Figure \ref{fig:cube_properties}). Therefore, it is not possible to intuitively predict whether the AQNs will generally be warmer (more emission) or colder (less emission). Figure \ref{fig:redshift_Inu} demonstrates that the decisive factor is primarily the ionization level of the environment: for a given redshift, i.e., with fixed number densities and baryon temperature, a more ionized environment leads to a higher AQN temperature and thus more AQN emission. The mixed environment shows a decrease in the emission at redshifts greater than $z\sim 5$. This is caused by the fact that the average baryon temperature decreases with redshift, and given our crude assumption of $T_{\rm cut}=2.6~\rm eV$ that separates ionized from neutral atomic states, the highest redshift cube turns out to be less ionized than all others. This decrease in intensity is also visible in Figure \ref{fig:Taqn_scatter_plot} as a decrease of the AQN temperature for $T_{\rm gas}< 2.6~\rm eV$. Additionally, Figure \ref{fig:redshift_Inu} indicates that in an ionized environment, large redshifts $z\gtrsim 2-3$ are the dominant contributors to the intensity measured at $z=0$. This implies that when searching for the AQN signature using cross-correlations, the use of high-redshift tracers would be our best chance.

\begin{figure}
    \begin{center}
	\includegraphics[width=13cm]{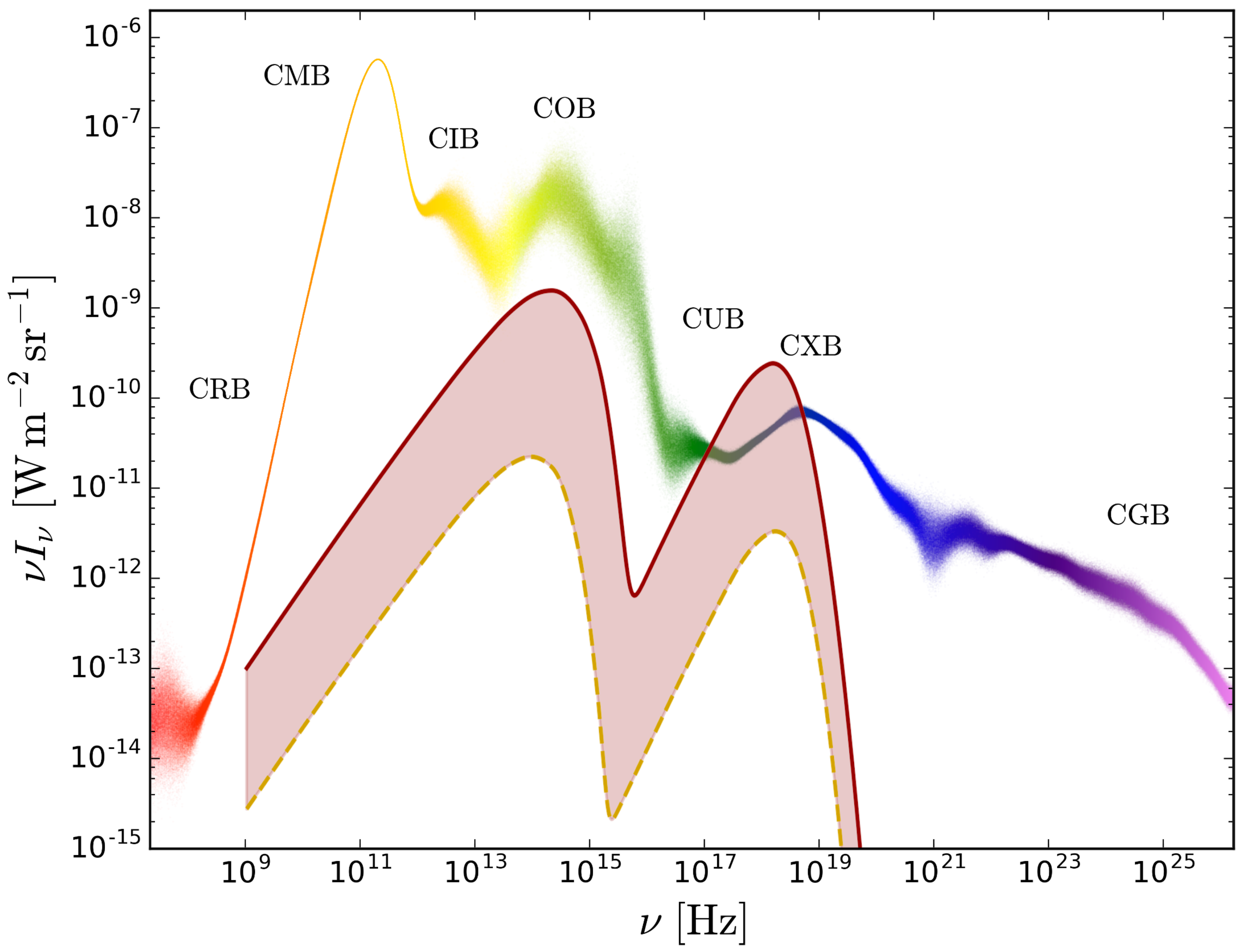}
	\end{center}
    \caption{The colored scattered points represent the overall estimate of the continuous cosmic background, consistent with observations. This is Figure 9 from \cite{2018ApSpe..72..663H}. The light red region corresponds to the AQN global emission, delimited by a maximum (top red solid line) and a minimum (bottom yellow dashed line) corresponding to ($m_\aqn=100~\rm g$, fully ionized) and ($m_\aqn=10~\rm g$, mixed). The neutral atomic case is not shown here, but would have a spectral shape similar to the red solid line, although five orders of magnitude smaller.}
    \label{fig:all_global_Inu}
\end{figure}

\begin{figure}
    \begin{center}
	\includegraphics[width=16cm]{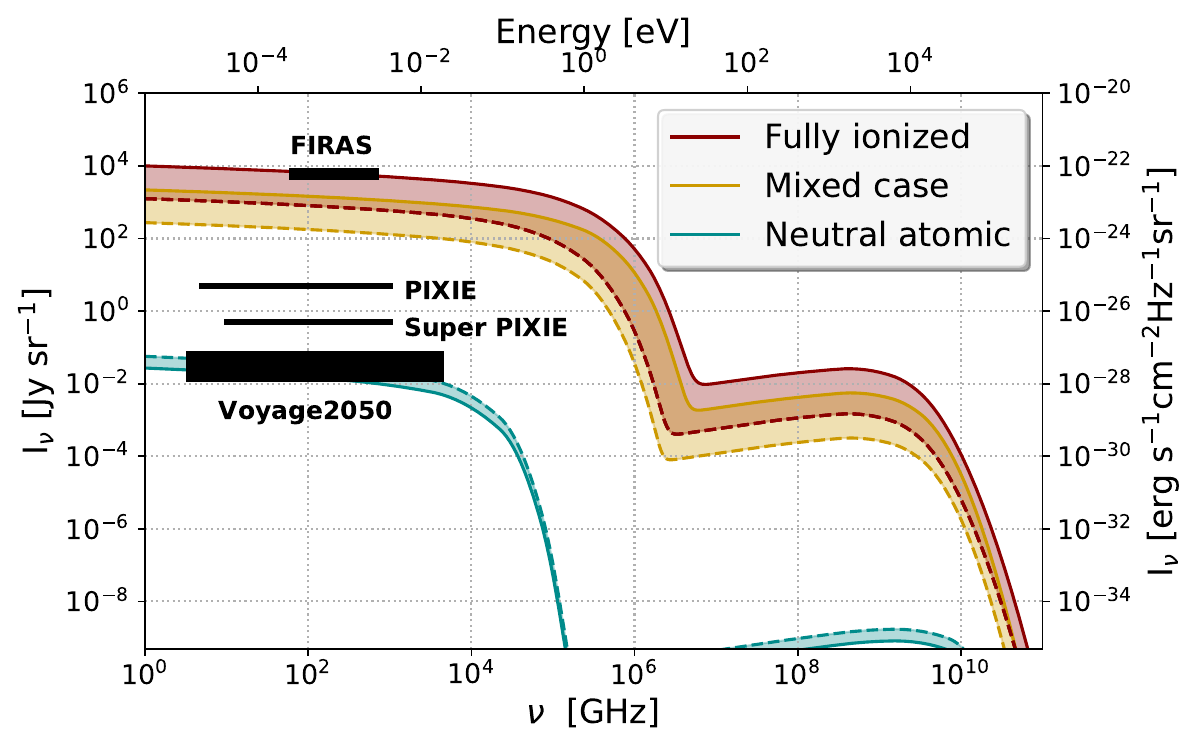}
	\end{center}
    \caption{Global intensity $I_\nu$ of the AQN signal for the three ionization cases and two AQN masses $m_\aqn = 10~\rm g$ (dashed line) and $m_\aqn = 100~\rm g$ (solid line). The FIRAS rectangle is the current precision of the CMB frequency spectrum. The black rectangles represent the sensitivity range of proposed experiments (references can be found in the text).}
    \label{fig:global_Inu}
\end{figure}

\begin{figure}
    \begin{center}
	\includegraphics[width=14cm]{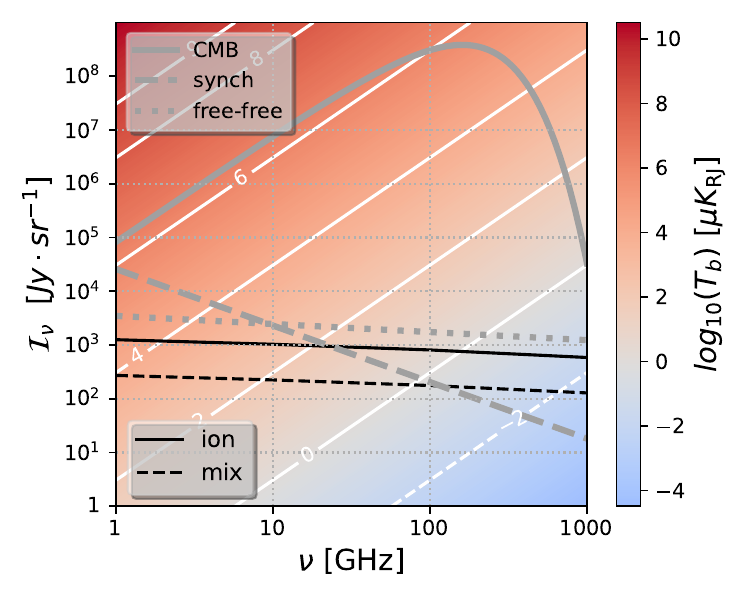}
	\end{center}
    \caption{Brightness temperature in $\mu{\rm K}$ assuming the Rayleigh-Jeans regime for a given intensity $I_\nu$ (white solid lines) and frequency $\nu$. The thick grey lines show the global specific intensity of the cosmic microwave background at $T_{\rm CMB}=2.725~\rm K$ (grey solid), synchrotron (grey dashed) and free-free (grey dotted) emissions. The black lines show the AQN emission for a mass $m_\aqn = 10~\rm g$ and two different ionized environments, fully ionized and mixed (solid and dashed black lines respectively).}
    \label{fig:radio_units}
\end{figure}

\section{Results}
\label{sec:results}

The AQN emission will generate a monopole signal (i.e. global intensity) and anisotropies (i.e. intensity fluctuations). Both will be computed using the light-cone simulations. In the following, the calculations are performed using a combination of three AQN masses $m_\aqn=[10,100,1000]~\rm g$ for three different environments: fully ionized, mixed and fully neutral atomic. The intensity of the AQN signal can vary by several orders of magnitude for the different mass/environment configurations. A most optimistic (maximum signal) and minimal configuration (minimal signal) will be identified below.

\subsection{Monopole}
The first step is to estimate the global AQN signal and compare it to the observed sky brightness as a function of frequency. In a recent review paper \citep{2018ApSpe..72..663H}, the average sky brightness observations across all wavelengths were compiled, from radio to $\gamma$-ray, which is precisely what the AQN signal prediction should be compared to. Figure 9 from \cite{2018ApSpe..72..663H} was copied in Figure \ref{fig:all_global_Inu} and overlaid with the AQN global intensity calculated from the previous sections. The AQN signal is presented as a region delimited by a maximum (top red solid line) and a minimum (bottom yellow dashed line) corresponding to ($m_\aqn=100~\rm g$, fully ionized) and ($m_\aqn=10~\rm g$, mixed). The neutral atomic case is not shown because the line would be four orders of magnitude below the yellow dashed line, and it is known to not correspond to the real Universe. The maximum signal (top red solid line on Figure \ref{fig:all_global_Inu}) and the neutral atomic case respectively define our most optimistic and minimum configurations.
The light red region on Figure \ref{fig:all_global_Inu} highlights what we believe is the most realistic range for the AQN model\footnote{Other combinations of AQN mass and environment will be discussed later.}. The two peaks at $\sim 10^{14}~\rm Hz$ and $\sim 10^{18}~\rm Hz$ correspond to the thermal and non-thermal emissions (see Figure \ref{fig:dFnu_tot}). Overall, the AQN signal is, up to a few orders of magnitude, below the measured sky brightness, but as we will see later, possibly within the detection limit of recently operational instruments and upcoming observatories. On Figure \ref{fig:all_global_Inu}, the maximum non-thermal emission exceeds the measured sky brightness in the soft X-ray regime, however, this should not be interpreted as ruling out the AQN model, for the following reasons: 1) depending on the AQN mass and the environment ionization, the non-thermal signal can be several orders of magnitude smaller; 2) As discussed in Section \ref{Fnu_nonth}, the calculations of the non-thermal emission are based on the oversimplified assumption that all positrons emit Bremsstralhung with the same cut-off frequency $\nu_c$ \citep{Forbes:2006ba}. In reality, positrons have a broad velocity distribution, resulting in a broader spectrum than in Figure \ref{fig:all_global_Inu}, but this is left for future work. Figure \ref{fig:global_Inu} shows the thermal and non-thermal signals together for the three ionization environments and the mass range $m_\aqn=[10-100]~\rm g$. In the ionized environment, the signal increases with $m_\aqn$, while it decreases in a neutral atomic environment. The latter is expected because the ratio AQN area over volume scales as $R_\aqn^{-1}$, meaning that for fixed total dark matter, more massive AQNs emit less radiation on average. On the other hand, in an ionized environment, this effect is compensated, and in fact reversed, by the increase of the effective radius. Consequently, in an ionized environment, the absence of an AQN signal detection will set an upper bound constraint on $m_\aqn$, while in a neutral atomic environment it will set a lower bound. The current detection limit of FIRAS, and of a few proposed experiments to measure the global intensity of the microwave sky, are shown on Figure \ref{fig:global_Inu}. Within, what is likely, the most realistic range for $m_\aqn$, and for a mixed or fully ionized environment, the AQN signal is not dramatically below the FIRAS detection limit. Experiments such as PIXIE \citep{2011JCAP...07..025K}, PRISM \citep{2013arXiv1306.2259P,2014JCAP...02..006A}, superPIXIE \citep{2019BAAS...51g.113K}, and Voyage2050 \citep{2021ExA....51.1515C} would have no problem detecting it. Interestingly, the sensitivity of Voyage2050 would even be capable of detecting the AQN signature in the most unlikely situation of a neutral atomic environment. The thermal emission (described in Section \ref{sec:thermal}) which is strong in the radio, up to the optical frequencies, is build on much more theoretically solid computational  scheme than the X-ray emission. For this reason, in the rest of this work, we will focus on the thermal emission.

Figure \ref{fig:radio_units} shows the AQN thermal emission in the frequency range where the cosmic microwave background dominates the cosmological signal. The synchrotron and free-free foregrounds are shown as indicators of where the AQN model predictions stand. The Galactic dust and the Cosmic Infrared Background are not shown; they scale as $\sim \nu^2$, cross the primary CMB signal at $\nu\sim 800~\rm GHz$ and are $\sim 10^4\,\rm Jy/sr$ at $\nu=100\,\rm GHz$. Other sources of spectral distortion, such as the $\mu$ and $y$ distortions, are shown on Figure 5 of \citep{2019BAAS...51g.113K}. Although the most optimistic AQN scenario is only marginally below the dominant foregrounds, the AQN signal of the minimal configuration (neutral atomic case) is comparable to the recombination signal of the CMB at $\sim 10^{-2}-10^{-1}\,\rm Jy/sr$, which is the ultimate target of proposed spectral distortion experiments. At first approximation, the AQN signal could be confused with a free-free spectrum, but the slope is slightly different and future work will have to demonstrate if they can be disentangled, in particular with components separation techniques combined with cross-correlations.

\subsection{Anisotropies}

\begin{figure}
 \noindent\makebox[\textwidth]{%
  \hspace{-10pt}%
  \includegraphics[width=1.2\textwidth]{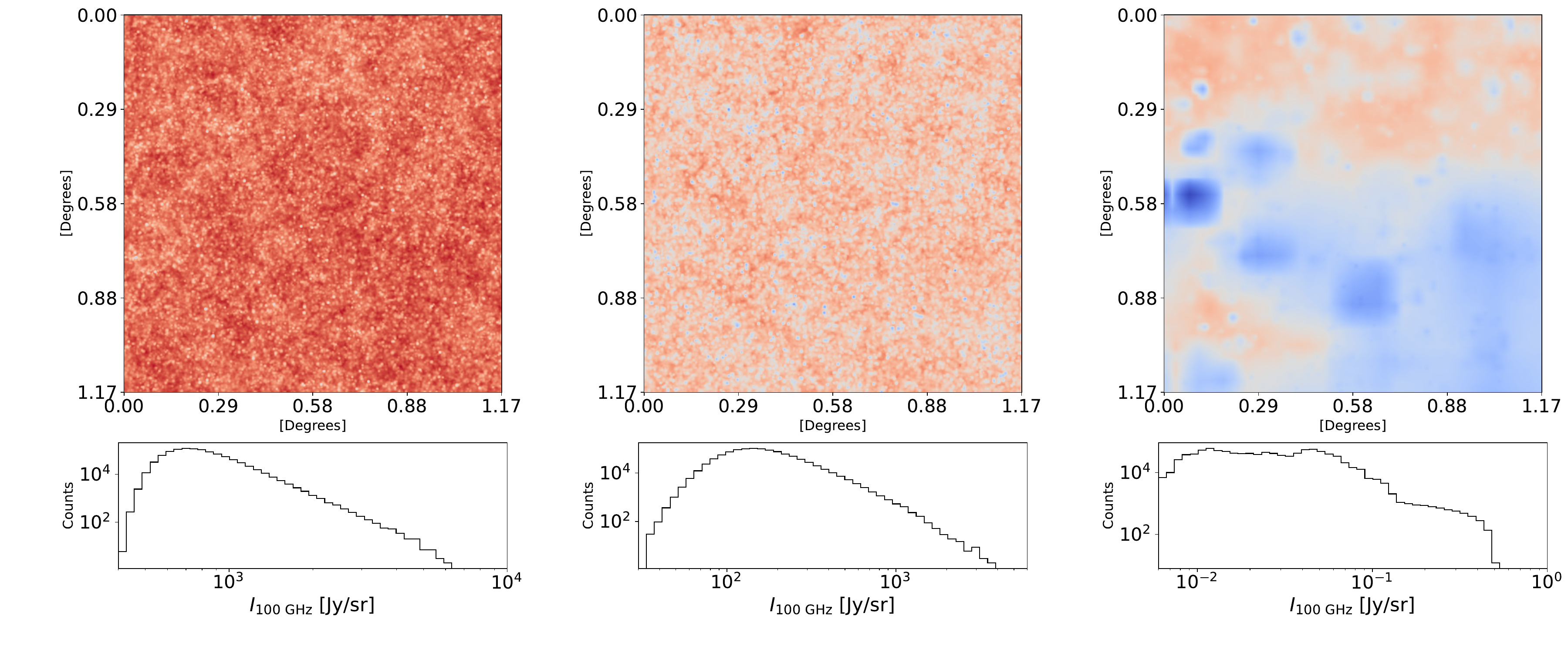}
}
    \caption{Specific intensity $I_\nu(\vec \theta)$ at $\nu=100~\rm GHz$ in units of Jy/sr for the three different environments, ionized, mix and neutral atomic for left, middle, right panels respectively. The top row shows the full light-cone $1.17\times 1.17$ sq. deg. and the bottom row shows the histogram of $I_\nu(\vec \theta)$.}
    \label{fig:light_cones}
\end{figure}

Intensity anisotropies is another prediction of the AQN model. Figure \ref{fig:light_cones} shows the intensity anisotropies in the radio sky ($\nu \lesssim 1000~\rm GHz$) for the three different ionization environments at $\nu = 100~\rm GHz$. The neutral atomic environment (right panel) shows large structures at low redshift, since the low redshift universe dominates the signal, as was already seen in Figure \ref{fig:redshift_Inu}. In the fully ionized case (left panel), it is the opposite, high redshift structures dominate and therefore exhibits very small structures, with a much larger intensity, also in agreement with Figure \ref{fig:redshift_Inu}. The histograms show that the r.m.s. of the anisotropies have the same order of magnitude as the average of the signal itself, regardless of the ionization environment. This is potentially worrying since the primary CMB anisotropies are $\sim 10^5$ smaller than the CMB monopole intensity, which would make them comparable to the AQN signal. Fortunately, Figure \ref{fig:light_cones} shows that when the intensity of the AQN radio anisotropies is high (ionized environment), the fluctuations take place at small angular scale, while when the anisotropy signal is low (neutral atomic environment), the fluctuations occur at relatively large angular scale (degree scale). 

\begin{figure}
    \begin{center}
	\includegraphics[width=13cm]{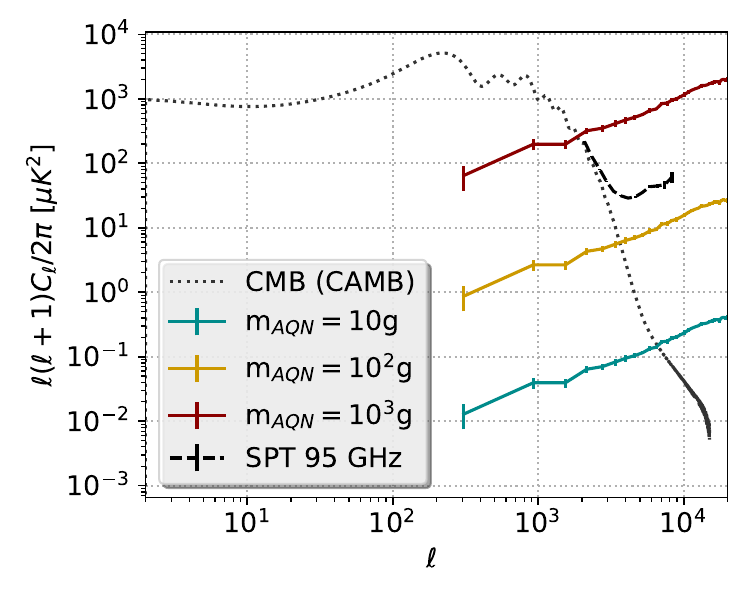}
	\end{center}
    \caption{Anisotropy angular power spectrum for the CMB from the South Pole Telescope experiment at 95 GHz (dashed line, from \citep{2012ApJ...755...70R}), compared to the theoretical spectrum using CAMB (dotted line). The colored solid lines show the AQN anisotropy signal at $\nu = 95~\rm GHz$ for different AQN masses in a fully ionized environment.}
    \label{fig:Cell_CMB_AQNs}
\end{figure}

One can use the AQN signal maps from Figure \ref{fig:light_cones} to calculate the angular power spectrum $C^\aqn_\ell$ \footnote{This is done using the \texttt{NaMaster} package available from \href{https://namaster.readthedocs.io/en/latest/}{\texttt{https://namaster.readthedocs.io/en/latest/}} \cite{2019MNRAS.484.4127A}} to compare it to the known radio sky anisotropies. Figure \ref{fig:Cell_CMB_AQNs} shows that for the fully ionized case, the high amplitude of anisotropies occur at high $\ell \sim 10^4$ \footnote{This is the limit of the Nbody simulation}, where the fluctuations of the primary CMB fluctuations are vanishing small and the signal is dominated by secondary anisotropies. Therefore, we can assert that the high amplitude of the AQN radio anisotropies do not conflict with the observed CMB anisotropy spectrum. The fully neutral atomic case is not shown, it peaks at $\ell \sim 200-400$ with much lower amplitude than the primary CMB (as expected from the right panel of Figure \ref{fig:light_cones}). 

How significantly are the AQN anisotropy contributing to the foreground secondary anisotropies? At $\ell \gtrsim 10^3$, the secondary anisotropies dominate the radio sky. For frequencies of interest in this work, $\nu \gtrsim 100~\rm GHz$, the two main contributors of the extragalactic foregrounds are the thermal Sunyaev-Zeldovich and the Cosmic Infrared Background. The black dashed line on Figure \ref{fig:Cell_CMB_AQNs} shows the South Pole Telescope signal for the 95 GHz channel \citep{2012ApJ...755...70R}, dominated by these foregrounds above $\ell \gtrsim 3000$. The AQN signal could be a contributor to the extragalactic foreground, but with an amplitude strongly dependent on $m_\aqn$. If one assumes the delta function distribution for the AQN masses (as described in Sect. \ref{sect:mass-distribution}) then figure \ref{fig:Cell_CMB_AQNs} naively suggests that high AQN mass ($m_\aqn \gtrsim 10^2~\rm g$) is apparently ruled out. However, any realistic mass distribution, e.g. a power law, is likely to decrease the intensity of the emission, which warrants a more careful and precise analysis of this region  in future. 

Since the AQN signal decreases as $\nu^{-2}$, it is expected that the extragalactic CIB becomes dominant at higher frequency $\nu > 100~\rm GHz$. The best frequency-scale window to look for the AQN anisotropy signal is $\ell \gtrsim 4000$ and $\nu \lesssim 100~\rm GHz$, although this is also where the synchrotron foregrounds are dominant. One should also keep in mind that the AQN anisotropy signal exists for the entire frequency spectrum displayed in Figure \ref{fig:global_Inu}. The detectability of the anisotropies in the infrared/optical spectrum is discussed in the section \ref{sect:future}.

\section{Discussion}
\label{sec:conclusion}

In this paper, we have explored the electromagnetic signatures of the AQN dark matter model at cosmological scales. The AQN framework, initially motivated by the similar values of the baryon and dark matter mass density parameters, originates from the standard model of particle physics and incorporates the QCD axion as a solution to the strong $\cal CP$ problem. Conventional dark matter models usually assume the existence of a new dark sector weakly or very weakly interacting with normal matter. AQNs, on the other hand, interact strongly with normal matter without violating the canons of the Big-Bang cosmology. Moreover, this exotic model may offer a solution to several outstanding cosmological problems. 

\subsection{Key results}

1- The thermal radiation is dominated by Bremsstrahlung and is determined by the electrosphere temperature $T_\aqn$. The main factor determining $T_\aqn$, and therefore the amplitude of the emission, is the ionization fraction of the baryonic environment, higher ionization leading to higher emission. The underlying physical mechanism is that AQNs with $T_\aqn\ne 0$ carry a small electrical charge, hence the Coulomb cross-section could significantly increase in an ionized environment.

2- The global intensity (monopole) is a signal which will modify the frequency spectrum of the combined synchrotron and free-free foregrounds. This is particularly relevant in the radio regime $\nu < 1000~\rm GHz$, where the primary CMB and the foreground monopoles are strong. The amplitude is a function of the AQN mass $m_\aqn$ and of the ionization level of the environment. Higher $m_\aqn$ leads to higher amplitude, therefore current measurement could provide an upper bound for the AQN mass. A realistic average AQN mass predicts an amplitude one to two orders of magnitude below the current FIRAS limits.

3- For a given frequency, the AQN anisotropy amplitude has an intensity comparable to the global signal. The reason is that the AQN signal originates from the low redshift universe, where mass density fluctuations are strongly non-linear. The angular power spectrum analysis of the anisotropies at $\nu=100~\rm GHz$ showed that the $C_\ell$ peaks at high $\ell\sim 10^4$ with a high amplitude in a fully ionized environment and much lower $\ell\sim 10^2$ with a low amplitude in a fully neutral atomic environment. The AQN $C_\ell$ is not in conflict with measurements: the high amplitude case occurs at very small angular scale where the AQN signal is much smaller than extragalactic foregrounds (thermal Sunyaev-Zeldovich and Cosmic Infrared Background) and the low amplitude case at large scale is much smaller than the primary CMB. 

4- The non-thermal radiation, which can extend up to a few dozens of keV, is caused by the positrons emitting Bremstralhung at the point of impact. It is based on simplified assumptions, which tend to overestimate the intensity of the signal at the characteristic frequency $\nu_c$. Therefore, our prediction using the light-cone simulation should be taken with caution in the X-ray regime, although the predicted AQN energy spectrum monopole is within the range of the current observational limits.

\subsection{Future work towards a possible detection}
\label{sect:future}

The light cone simulation reveals several interesting features which can help to design a target search for the AQN signal. The detection of the global signal and anisotropies requires distinct strategies.
\bigskip

1- Figures \ref{fig:all_global_Inu} and \ref{fig:global_Inu} show that the AQN thermal emission extends up to optical wavelength. It is the frequency range $\nu=[10^{13}-10^{15}]~\rm Hz$, where the intensity is the closest to the measured sky background, that the chance of a direct detection is highest. Table \ref{tab:optical_range} shows the expected surface brightness $I_\nu$ calculated from the monopole AQN intensity for the maximum ($m_\aqn = 100~\rm g$ and fully ionized) and minimum $m_\aqn=10~\rm g$ for the mixed ionized environment) cases shown in Figure \ref{fig:all_global_Inu}. We emphasize that, since the monopole and the anisotropies have similar intensity, they also have similar surface brightness, therefore the monopole $I_\nu$ is a good estimate, to first approximation, of the anisotropy $\delta I_\nu$.

Due to the Zodiacal light, the monopole is very difficult to measure in the infrared/optical. There are reports of Extragalactic Background Light excess across the infrared and optical spectrum (see the Figure 8 in \cite{2019ConPh..60...23M} and references therein). Using the New Horizons-LORRI data, the Cosmic Optical Background appears to be brighter than expected \citep{2022ApJ...927L...8L,2023ApJ...945...45S} and it is debated whether it could be linked to dark matter \citep{2022PhRvL.129w1301B,2022PhRvD.106j3505N}. However, constraints on AQN in the infrared/optical are most likely to first come from anisotropies. Anisotropy measurements already exist \citep{2015NatCo...6.7945M}, and very low surface brightness fluctuations, such as shown in Table \ref{tab:optical_range}, are in principle achievable with the new generation of space based telescopes, Euclid \citep{2024arXiv240513496C} and the James Webb Space Telescope. This is left for future work, but ultimately, the consistency between the monopole and the anisotropy signals would constitute a strong test in the infrared/optical range (e.g. \cite{2022PhRvD.106j3505N,2024MNRAS.528.5019H}).

\begin{table}[h]
    \centering
    \begin{tabular}{|c|c|c|}
        \hline
        $\nu$ [Hz] & $\lambda$ [$\mu$m] & $\left. I_\nu\right|_{\rm max,min}$ [$m_{\rm AB}$/arcsec$^2$] \\
        \hline
        $10^{13}$ & 30 & 26.7, 30.7  \\
        \hline
        $10^{14}$ & 3 & 27.6, 32.1  \\
        \hline
        $3\times 10^{14}$ & 1.0 & 28.7,   33.8  \\
        \hline
        $4.3\times 10^{14}$ & 0.7 & 29.3,   35.1  \\
        \hline
        $10^{15}$ & 0.3 & 31.2,   39.0  \\
        \hline
    \end{tabular}
    \caption{Monopole surface brightness of the AQN signal in the infrared and optical range. $\left. I_\nu\right|_{\rm max}$ and $\left. I_\nu\right|_{\rm min}$ are taken from the light red region of Figure \ref{fig:all_global_Inu} for the corresponding frequencies.}
    \label{tab:optical_range}
\end{table}

2- Another interesting path of investigation is suggested by Figure \ref{fig:Cell_CMB_AQNs}: the South Pole Telescope will be reanalysed with the AQN signal amplitude parametrized as a function of $\ell$ and $\nu$. This should give a strong upper bound on the AQN mass $m_\aqn$. Extending this analysis to much lower frequency, $1~{\rm GHz} \lesssim \nu \lesssim 100~{\rm GHz}$, would be particularly interesting, since the AQN emission amplitude should scale as $\sim \nu^{-2}$ due to the near Bremsstrahlung-like spectral energy distribution. The Very Large Array Sky Survey, which is operating at $2-4~\rm GHz$ with a nominal sensitivity of $69~\rm \mu Jy$, seems to be particularly well suited \citep{2020PASP..132c5001L}. Figure \ref{fig:radio_units} also suggests another interesting test, by letting the AQN mass vary and using components separation to extract the primary CMB and the foreground signals.

3- Figure \ref{fig:redshift_Inu} shows that most of the global and fluctuation signals is coming from redshift $z>1-2$. A direct AQN signal search, such as proposed in the previous two paragraphs could be improved if we cross-correlate data sets with a chosen redshift range overlap, such that the resulting signal-to-noise is improved while reducing contaminating sources, e.g. for sources at $z<1$. A proof of concept that cross-correlations can boost the sensitivity and target a specific population was performed in \cite{2023MNRAS.525.1443L}. By cross-correlating Herschel-SPIRE imaging and the Canada-France Imaging Survey (CFIS) Low Surface Brightness data \footnote{Products from the Ultraviolet Near-Infrared Optical Northern Survey (UNIONS)}, it was shown that an r.m.s. surface brightness of $r_{\rm AB}\sim 32.5~\rm mag/arcsec^2$ is achievable statistically. According to Table \ref{tab:optical_range}, this would be in the range of interest.

4- Another approach consists in looking for the AQN signature in specific environments. In this work we focused on large scale structures, but the physical parameters characterizing the baryonic environment, such as gas temperature, ionization level and baryon density, are greatly dependent on the system. The AQN signature in radio, optical, UV and X-ray is being investigated in clusters of galaxies \citep{2024arXiv240617946S} and the Milky Way \citep{sekatchev2024prep}. A different environment is not necessarily a specific object. For instance, we are investigating the effect of AQN energy injection in the pre-recombination era of the CMB, specifically how the $\mu$ and $y$ spectral distortions and the optical depth are changed \citep{majidi2024prep}.

\subsection{Future work: theory}

%Several important aspects of the AQN framework are still under development, such as a more realistic treatment of the non-thermal emission. 
A particularly interesting regime where AQN physics could be relevant is during the Dark Ages and the Cosmic Dawn. Although the Universe is in atomic form at the $\sim 99.9\%$ level during the dark ages, the number density of matter scales as $(1+z)^3$, significantly increasing the annihilation rate. The resulting energy injection between recombination and reionization will precede the formation of the first stars, changing the number density of free electrons and, consequently, the optical depth. Constraints on energy injection during the dark ages already exist for decaying and annihilating dark matter models, primarily through the optical depth measured from Planck \cite{2020A&A...641A...6P,2019JCAP...04..025G}. AQNs can also alter the history of recombination. The challenge with these calculations is that, unlike dark matter decay and annihilation models, the injected photons cover a large energy range, including the resonant Ly$\alpha$ line. Therefore, proper inclusion of AQN physics in recombination and dark ages studies must account for atomic physics, not just the injected energy.

As the universe approaches cosmic dawn, a 21 cm line signal can develop, which the AQN physics can also impact. It is known that whether the 21 cm line is in emission or absorption at a given redshift $z$ depends on the temperature of the surrounding radiation $T_\gamma(z)$, the matter temperature $T_{\rm gas}(z)$, and the color temperature of the Ly$\alpha$ photons $T_\alpha(z)$. It is usually assumed that $T_\gamma$ is the CMB temperature at $z$, but if dark matter is made of AQNs, the impact on 21 cm line could be altered in two opposite ways: injected photons will provide an additional source of photons in the radio background, strengthening the 21 cm absorption, but high-energy photons will reheat the matter, reducing 21 cm absorption. The effect is similar to a decaying or annihilating dark matter \cite{2020JCAP...07..020S}, but with the AQN dark matter a full atomic treatment is necessary.

The significant source of radio photons from AQN-baryon interaction can also have a cosmological consequence at low frequencies. Below $\nu < 1-10~\rm GHz$, an excess of radiation relative to the CMB, once the known foregrounds have been removed, has been known since the 80s, known as the ARCADE excess \cite{2011ApJ...734....5F}. Its origin, whether extragalactic or within the Milky Way, is still debated. The extrapolation of the AQN signal well below $\sim 1~\rm GHz$ in Figure \ref{fig:all_global_Inu} could be such a source, but at these frequencies, it is necessary to consider the physics of AQNs up to very high redshifts, possibly recombination, which requires significant theoretical development.
%The controversy that followed the first detection of 21 cm absorption \citep{2018Natur.555...67B}, twice as strong as expected, triggered a lot of suggestions that it could be associated with new physics. The solution has to either cool the gas, or heat the CMB.

\subsection{Concluding remarks}

The key element of the AQN construction is that the DM and visible components are described as two different phases of the same fundamental QCD theory of the Standard Model.
This model predicts the injection of energy to surrounding space at all redshifts and in very broad frequency bands, as we have shown in this work. The AQN signal is proportional to the dark and visible components, i.e. $\propto \rho_{\rm DM}\cdot n_{\rm b}$ \footnote{As opposed to $\propto \rho_{\rm DM}$ for decaying dark matter, and $\rho_{\rm DM}^2$ for annihilating dark matter}, but we have also shown that purely ordinary matter parameters such as the gas temperature and the ionization rate strongly impact the dark matter signature. Therefore, the state of the baryonic environment plays a crucial role in the intensity of the dark matter signal. This is the most dramatic difference between canonical DM models to the AQN framework because dark matter and baryonic matter are strongly and non-trivially entangled. This property may dramatically modify many features during the long evolution of the Universe.  The model is consistent with all presently available cosmological, astrophysical, satellite and ground-based observations and may even shed some light on a number of long-standing puzzles and mysteries as mentioned in the text. Without anticipating on the possible success or failure of the AQN model, it definitely offers an exciting and refreshing approach to the problem of the nature of dark matter, considering that the WIMP paradigm, despite a dramatic improvement of instrumentation over the past 40 years, still has not found any sign of dark matter.

\acknowledgments

FM, XL, LVW, AZ and MS acknowledges the support from NSERC. We thank Gary Hinshaw, Kirit Karkare and Ryley Hill for useful discussions. KD acknowledges support by the COMPLEX project from the European Research Council (ERC) under the European Union’s Horizon 2020 research and innovation program grant agreement ERC-2019-AdG 882679 as well as by the Deutsche Forschungsgemeinschaft (DFG, German Research Foundation) under Germany’s Excellence Strategy - EXC-2094 - 390783311.

The calculations for the hydrodynamical simulations were carried out at the 
Leibniz Supercomputer Center (LRZ) under the project pr83li.

The data in this paper were analyzed with open-source python packages \texttt{numpy \cite{2020Natur.585..357H}, scipy \cite{2020NatMe..17..261V}, astropy \cite{2018AJ....156..123A}, matplotlib \cite{2007CSE.....9...90H}}, and \texttt{NaMaster \cite{2019MNRAS.484.4127A}}.

%\paragraph{Note added.} This is also a good position for notes added
%after the paper has been written.

\appendix
\section{Natural and C.G.S. units conversion}
\label{app: units}

The natural unit expressions given in this paper are convenient in order to grasp the orders of magnitude of the different phenomena using well known particle physics quantities, however, a conversion to physical units is necessary in order to do the calculations. In this Appendix, we provide the recipe to convert various expressions between different unit system. Table \ref{tab:unit conversion} shows how to convert between C.G.S and natural unit for most quantities used in this work. Note that, with natural unit, any physical quantity has the dimension of energy, where the dimension could be any integer $\mathbb{Z}$. Energy is conveniently expressed in electron-volts, although Table \ref{tab:unit conversion} shows the unit conversion from/to natural units using the erg unit for energy.
In order to avoid any confusion, we should ideally indicate which units are used for a particular quantity, for instance, using Table \ref{tab:unit conversion} and $1~{\rm eV}=1.6021766\times 10^{-12}~\rm erg$, a length $\ell$ in C.G.S, natural (erg) and natural (eV) units would be given by:

\begin{equation}
    \ell_{\rm CGS}=\ell_{\rm Nat,erg^{-1}}(\hbar c)_{\rm CGS}=\ell_{\rm Nat,GeV^{-1}}\frac{(\hbar c)_{\rm CGS}}{0.0016021766}
\end{equation}
Implying that $1~\rm GeV^{-1}=1.9732698\times 10^{-14}~\rm cm$. Using the same reasoning, a time $t$ would be given by:

\begin{equation}
    t_{\rm CGS}=t_{\rm Nat,erg^{-1}}\hbar_{\rm CGS}=t_{\rm Nat,GeV^{-1}}\frac{\hbar_{\rm CGS}}{0.0016021766}
\end{equation}
implying that $1~\rm GeV^{-1}=6.5821196\times 10^{-25}~ s$. However, the choice of a particular unit system is generally not indicated, meaning that, when converting a particular quantity (e.g. flux) from natural to C.G.S, it is necessary to know that we are dealing with a flux (in $\rm erg~s^{-1}~cm^{-2}$).

\begin{table}[htbp]
\renewcommand{\arraystretch}{1.5}
	\centering
    \caption{Conversion table between C.G.S and natural units for most quantities used in this work, where the latter are expressed in erg, the unit of energy in C.G.S..}
	\begin{tabular}{l c c l} 
		\hline\hline
		Quantity & C.G.S & Natural [erg] \\ 
		\hline
		Length $\ell $& [cm]  & [erg $^{-1}$] = $\frac{\rm [cm]}{\hbar c}$  \\ 
		Time $t$ & [s]  & [erg $^{-1}$] = $\frac{\rm [s]}{\hbar}$   \\ 
		Mass $m$ & [g] & [erg] = [g]c$^2$  \\ 
		Velocity $\rm v$ & [cm s$^{-1}$] & [1] = [cm s$^{-1}$]$\frac{1}{c}$\\ 
        Momentum $\rm p$ & [g cm s$^{-1}$] & [erg] = [g cm s$^{-1}$]c\\
        Cross-section $\sigma$ & [cm$^2$] & [erg$^{-2}$] = $\frac{\rm [cm^2]}{(\hbar c)^2}$\\
        Collision rate $\Gamma $ & [s$^{-1}$]  & [erg] = [s$^{-1}]\hbar$  \\ 
        Frequency $\nu $ & [Hz]  & [erg] = [Hz]$h$  \\ 
        Angular frequency $2\pi\nu$ & [rad Hz] & [erg] = [rad Hz]$\hbar$ \\
        Temperature $T$ & [K]  & [erg] = [K]$k_{\rm B}$  \\ 

		\hline\hline
	\end{tabular}
	\label{tab:unit conversion}
\end{table}

\begin{table*}[htbp]
\renewcommand{\arraystretch}{1.5}
	\centering
    \caption{The equations referenced in the left column are written in natural and C.G.S units for completeness and to ease the numerical calculations.}
	\begin{tabular}{l c c l} 
		\hline\hline
		Eq.(\#) & Natural & C.G.S  \\ 
		\hline
 (\ref{eq:dFtot}) & $\frac{16}{3}T_\aqn^4\frac{\alpha^{5/2}}{\pi}\left(\frac{T_\aqn}{m_e}\right)^{1/4}$ & $\frac{16\alpha^{5/2}k_{\rm B}^4T_\aqn^4}{3\pi\hbar^3c^2}\sqrt[4]{\frac{k_{\rm B}T_\aqn}{m_e c^2}}$ \\
 
 (\ref{eq:Q}) & $\frac{4\pi R^2}{\sqrt{2\pi\alpha}}(m_e T_\aqn)\left(\frac{T_\aqn}{m_e}\right)^{1/4}$ & $\frac{4\pi R^2 k_{\rm B}m_e T_\aqn}{\sqrt{2\pi \alpha}\hbar^2}
\left(\frac{k_{\rm B}T_\aqn}{m_ec^2}\right)^{\frac{1}{4}}$ \\

(\ref{eq:Reff}) & $8\alpha m_e^2 R_\aqn^2 \left(\frac{T_\aqn}{T_{\rm gas}}\right)^2\sqrt{\frac{T_\aqn}{m_e}}$ & $8\alpha\left(\frac{m_e c R_\aqn}{\hbar}\right)^2
\left(\frac{T_\aqn}{T_{\rm gas}}\right)^2\sqrt{\frac{k_{\rm B}T_\aqn}{m_e c^2}}$ \\

(\ref{eq:Taqn_with_Reff}) & $m_e\left[\frac{2{\rm GeV}~f(1-g)}{8\alpha^{3/2}T_{\rm gas}^2}3\pi n_{\rm b} \Delta{\rm v} R_\aqn^2\right]^\frac{4}{7}$ & $\frac{m_ec^2}{k_{\rm B}}\left[\frac{3\pi}{8}\frac{2{\rm\,GeV}\,\hbar}{\alpha^{3/2}k_{\rm B}^2}\frac{f(1-g)R_\aqn^2 \Delta{\rm v} n_{\rm b}}{T_{\rm gas}^2}\right]^{\frac{4}{7}}$\\

(\ref{eq:Taqn_with_Rgeo}) & $m_e^{1/17}\left[\frac{3\pi}{4}\frac{2~{\rm GeV}}{16\alpha^{5/2}}f(1-g)\Delta{\rm v}~n_{\rm b}\right]^{4/17}$ & $\frac{1}{k_{\rm B}}\left[\frac{3\pi}{64}\frac{2{\rm\,GeV}\,\hbar^3c^{5/2}}{\alpha^{5/2}}f(1-g)m_e^{1/4}n_{\rm b} \Delta{\rm v}\right]^{\frac{4}{17}}$ \\

(\ref{eq:dFnu}) & $\frac{4}{45}\frac{T_\aqn^3\alpha^{5/2}}{\pi}\left(\frac{T_\aqn}{m_e}\right)^{1/4} H\left(\frac{h\nu}{T_\aqn}\right)$ & $\frac{4\alpha^{5/2}k_{\rm B}^3}{45\pi\hbar^2c^2}T_\aqn^3
\sqrt[4]{\frac{k_{\rm B}T_\aqn}{m_ec^2}}\,
H\left(\frac{h\nu}{k_{\rm B}T_\aqn}\right)$ \\

\eqref{eq:dFnu_xray ::2} & $\frac{16 \alpha^{5/2}}{3 \pi \Gamma(4/3)}\frac{g}{1-g}\frac{1}{\nu_c}\frac{T_\aqn^{17/4}}{m_e^{1/4}} \left(\frac{\nu}{\nu_c}\right)^{\frac{1}{3}}e^{-\nu/\nu_c}$ & $\frac{16 \alpha^{5/2}}{3 \pi \Gamma(4/3)}
\frac{k_{\rm B}^{17/4}}{\hbar^3c^{5/2}}
\frac{g}{1-g}\frac{1}{\nu_c}\frac{T_\aqn^{17/4}}{m_e^{1/4}} \left(\frac{\nu}{\nu_c}\right)^{\frac{1}{3}}e^{-\nu/\nu_c}$ \\

		\hline\hline
	\end{tabular}
	\label{tab:Eqs_cgs2nat}
\end{table*}

\bibliographystyle{JHEP}
\bibliography{JCAP_aqn_LSS_radio_revised.bib}

\providecommand{\href}[2]{#2}\begingroup\raggedright\begin{thebibliography}{100}

\bibitem{2020A&A...641A...6P}
{Planck Collaboration}, N.~{Aghanim}, Y.~{Akrami}, M.~{Ashdown}, J.~{Aumont},
  C.~{Baccigalupi} et~al., \emph{{Planck 2018 results. VI. Cosmological
  parameters}}, \href{https://doi.org/10.1051/0004-6361/201833910}{\emph{\aap}
  {\bfseries 641} (2020) A6}
  [\href{https://arxiv.org/abs/1807.06209}{{\ttfamily 1807.06209}}].

\bibitem{2018Natur.562...51B}
G.~{Bertone} and T.M.P.~{Tait}, \emph{{A new era in the search for dark
  matter}}, \href{https://doi.org/10.1038/s41586-018-0542-z}{\emph{\nat}
  {\bfseries 562} (2018) 51}
  [\href{https://arxiv.org/abs/1810.01668}{{\ttfamily 1810.01668}}].

\bibitem{Mohapatra:1998nd}
R.N.~Mohapatra and V.L.~Teplitz, \emph{{Primordial nucleosynthesis constraint
  on massive, stable, strongly interacting particles}},
  \href{https://doi.org/10.1103/PhysRevLett.81.3079}{\emph{Phys. Rev. Lett.}
  {\bfseries 81} (1998) 3079}
  [\href{https://arxiv.org/abs/hep-ph/9804420}{{\ttfamily hep-ph/9804420}}].

\bibitem{Cyburt:2002uw}
R.H.~Cyburt, B.D.~Fields, V.~Pavlidou and B.D.~Wandelt, \emph{{Constraining
  strong baryon dark matter interactions with primordial nucleosynthesis and
  cosmic rays}}, \href{https://doi.org/10.1103/PhysRevD.65.123503}{\emph{Phys.
  Rev. D} {\bfseries 65} (2002) 123503}
  [\href{https://arxiv.org/abs/astro-ph/0203240}{{\ttfamily
  astro-ph/0203240}}].

\bibitem{Ali-Haimoud:2015pwa}
Y.~Ali-Ha\"\i{}moud, J.~Chluba and M.~Kamionkowski, \emph{{Constraints on Dark
  Matter Interactions with Standard Model Particles from Cosmic Microwave
  Background Spectral Distortions}},
  \href{https://doi.org/10.1103/PhysRevLett.115.071304}{\emph{Phys. Rev. Lett.}
  {\bfseries 115} (2015) 071304}
  [\href{https://arxiv.org/abs/1506.04745}{{\ttfamily 1506.04745}}].

\bibitem{Ali-Haimoud:2021lka}
Y.~Ali-Ha\"\i{}moud, \emph{{Testing dark matter interactions with CMB spectral
  distortions}}, \href{https://doi.org/10.1103/PhysRevD.103.043541}{\emph{Phys.
  Rev. D} {\bfseries 103} (2021) 043541}
  [\href{https://arxiv.org/abs/2101.04070}{{\ttfamily 2101.04070}}].

\bibitem{Dubovsky:2001tr}
S.L.~Dubovsky and D.S.~Gorbunov, \emph{{Small second acoustic peak from
  interacting cold dark matter?}},
  \href{https://doi.org/10.1103/PhysRevD.64.123503}{\emph{Phys. Rev. D}
  {\bfseries 64} (2001) 123503}
  [\href{https://arxiv.org/abs/astro-ph/0103122}{{\ttfamily
  astro-ph/0103122}}].

\bibitem{Chen:2002yh}
X.~{Chen}, S.~{Hannestad} and R.J.~{Scherrer}, \emph{{Cosmic microwave
  background and large scale structure limits on the interaction between dark
  matter and baryons}},
  \href{https://doi.org/10.1103/PhysRevD.65.123515}{\emph{\prd} {\bfseries 65}
  (2002) 123515}.

\bibitem{Dvorkin:2013cea}
C.~{Dvorkin}, K.~{Blum} and M.~{Kamionkowski}, \emph{{Constraining dark
  matter-baryon scattering with linear cosmology}},
  \href{https://doi.org/10.1103/PhysRevD.89.023519}{\emph{\prd} {\bfseries 89}
  (2014) 023519} [\href{https://arxiv.org/abs/1311.2937}{{\ttfamily
  1311.2937}}].

\bibitem{Gluscevic:2017ywp}
V.~Gluscevic and K.K.~Boddy, \emph{{Constraints on Scattering of
  keV\textendash{}TeV Dark Matter with Protons in the Early Universe}},
  \href{https://doi.org/10.1103/PhysRevLett.121.081301}{\emph{Phys. Rev. Lett.}
  {\bfseries 121} (2018) 081301}
  [\href{https://arxiv.org/abs/1712.07133}{{\ttfamily 1712.07133}}].

\bibitem{Boddy:2018kfv}
K.K.~Boddy and V.~Gluscevic, \emph{{First Cosmological Constraint on the
  Effective Theory of Dark Matter-Proton Interactions}},
  \href{https://doi.org/10.1103/PhysRevD.98.083510}{\emph{Phys. Rev. D}
  {\bfseries 98} (2018) 083510}
  [\href{https://arxiv.org/abs/1801.08609}{{\ttfamily 1801.08609}}].

\bibitem{Boddy:2018wzy}
K.K.~Boddy, V.~Gluscevic, V.~Poulin, E.D.~Kovetz, M.~Kamionkowski and
  R.~Barkana, \emph{{Critical assessment of CMB limits on dark matter-baryon
  scattering: New treatment of the relative bulk velocity}},
  \href{https://doi.org/10.1103/PhysRevD.98.123506}{\emph{Phys. Rev. D}
  {\bfseries 98} (2018) 123506}
  [\href{https://arxiv.org/abs/1808.00001}{{\ttfamily 1808.00001}}].

\bibitem{Xu:2018efh}
W.L.~Xu, C.~Dvorkin and A.~Chael, \emph{{Probing sub-GeV Dark Matter-Baryon
  Scattering with Cosmological Observables}},
  \href{https://doi.org/10.1103/PhysRevD.97.103530}{\emph{Phys. Rev. D}
  {\bfseries 97} (2018) 103530}
  [\href{https://arxiv.org/abs/1802.06788}{{\ttfamily 1802.06788}}].

\bibitem{Li:2018zdm}
Z.~Li, V.~Gluscevic, K.K.~Boddy and M.S.~Madhavacheril, \emph{{Disentangling
  Dark Physics with Cosmic Microwave Background Experiments}},
  \href{https://doi.org/10.1103/PhysRevD.98.123524}{\emph{Phys. Rev. D}
  {\bfseries 98} (2018) 123524}
  [\href{https://arxiv.org/abs/1806.10165}{{\ttfamily 1806.10165}}].

\bibitem{Li:2022mdj}
Z.~Li et~al., \emph{{The Atacama Cosmology Telescope: limits on dark
  matter-baryon interactions from DR4 power spectra}},
  \href{https://doi.org/10.1088/1475-7516/2023/02/046}{\emph{JCAP} {\bfseries
  02} (2023) 046} [\href{https://arxiv.org/abs/2208.08985}{{\ttfamily
  2208.08985}}].

\bibitem{Tashiro:2014tsa}
H.~Tashiro, K.~Kadota and J.~Silk, \emph{{Effects of dark matter-baryon
  scattering on redshifted 21 cm signals}},
  \href{https://doi.org/10.1103/PhysRevD.90.083522}{\emph{Phys. Rev. D}
  {\bfseries 90} (2014) 083522}
  [\href{https://arxiv.org/abs/1408.2571}{{\ttfamily 1408.2571}}].

\bibitem{Munoz:2015bca}
J.B.~Mu\~noz, E.D.~Kovetz and Y.~Ali-Ha\"\i{}moud, \emph{{Heating of Baryons
  due to Scattering with Dark Matter During the Dark Ages}},
  \href{https://doi.org/10.1103/PhysRevD.92.083528}{\emph{Phys. Rev. D}
  {\bfseries 92} (2015) 083528}
  [\href{https://arxiv.org/abs/1509.00029}{{\ttfamily 1509.00029}}].

\bibitem{Munoz:2017qpy}
J.B.~Mu\~noz and A.~Loeb, \emph{{Constraints on Dark Matter-Baryon Scattering
  from the Temperature Evolution of the Intergalactic Medium}},
  \href{https://doi.org/10.1088/1475-7516/2017/11/043}{\emph{JCAP} {\bfseries
  11} (2017) 043} [\href{https://arxiv.org/abs/1708.08923}{{\ttfamily
  1708.08923}}].

\bibitem{Rogers:2021byl}
K.K.~Rogers, C.~Dvorkin and H.V.~Peiris, \emph{{Limits on the Light Dark
  Matter\textendash{}Proton Cross Section from Cosmic Large-Scale Structure}},
  \href{https://doi.org/10.1103/PhysRevLett.128.171301}{\emph{Phys. Rev. Lett.}
  {\bfseries 128} (2022) 171301}
  [\href{https://arxiv.org/abs/2111.10386}{{\ttfamily 2111.10386}}].

\bibitem{Nadler:2019zrb}
E.O.~Nadler, V.~Gluscevic, K.K.~Boddy and R.H.~Wechsler, \emph{{Constraints on
  Dark Matter Microphysics from the Milky Way Satellite Population}},
  \href{https://doi.org/10.3847/2041-8213/ab1eb2}{\emph{Astrophys. J. Lett.}
  {\bfseries 878} (2019) 32}
  [\href{https://arxiv.org/abs/1904.10000}{{\ttfamily 1904.10000}}].

\bibitem{DES:2020fxi}
{\scshape DES} collaboration, \emph{{Milky Way Satellite Census. III.
  Constraints on Dark Matter Properties from Observations of Milky Way
  Satellite Galaxies}},
  \href{https://doi.org/10.1103/PhysRevLett.126.091101}{\emph{Phys. Rev. Lett.}
  {\bfseries 126} (2021) 091101}
  [\href{https://arxiv.org/abs/2008.00022}{{\ttfamily 2008.00022}}].

\bibitem{Maamari:2020aqz}
K.~Maamari, V.~Gluscevic, K.K.~Boddy, E.O.~Nadler and R.H.~Wechsler,
  \emph{{Bounds on velocity-dependent dark matter-proton scattering from Milky
  Way satellite abundance}},
  \href{https://doi.org/10.3847/2041-8213/abd807}{\emph{Astrophys. J. Lett.}
  {\bfseries 907} (2021) L46}
  [\href{https://arxiv.org/abs/2010.02936}{{\ttfamily 2010.02936}}].

\bibitem{Chivukula:1989cc}
R.S.~Chivukula, A.G.~Cohen, S.~Dimopoulos and T.P.~Walker, \emph{{Bounds on
  Halo Particle Interactions From Interstellar Calorimetry}},
  \href{https://doi.org/10.1103/PhysRevLett.65.957}{\emph{Phys. Rev. Lett.}
  {\bfseries 65} (1990) 957}.

\bibitem{Bhoonah:2018wmw}
A.~Bhoonah, J.~Bramante, F.~Elahi and S.~Schon, \emph{{Calorimetric Dark Matter
  Detection With Galactic Center Gas Clouds}},
  \href{https://doi.org/10.1103/PhysRevLett.121.131101}{\emph{Phys. Rev. Lett.}
  {\bfseries 121} (2018) 131101}
  [\href{https://arxiv.org/abs/1806.06857}{{\ttfamily 1806.06857}}].

\bibitem{Wadekar:2019mpc}
D.~Wadekar and G.R.~Farrar, \emph{{Gas-rich dwarf galaxies as a new probe of
  dark matter interactions with ordinary matter}},
  \href{https://doi.org/10.1103/PhysRevD.103.123028}{\emph{Phys. Rev. D}
  {\bfseries 103} (2021) 123028}
  [\href{https://arxiv.org/abs/1903.12190}{{\ttfamily 1903.12190}}].

\bibitem{Bhoonah:2020dzs}
A.~Bhoonah, J.~Bramante, S.~Schon and N.~Song, \emph{{Detecting composite dark
  matter with long-range and contact interactions in gas clouds}},
  \href{https://doi.org/10.1103/PhysRevD.103.123026}{\emph{Phys. Rev. D}
  {\bfseries 103} (2021) 123026}
  [\href{https://arxiv.org/abs/2010.07240}{{\ttfamily 2010.07240}}].

\bibitem{Qin:2001hh}
B.~Qin and X.-P.~Wu, \emph{{Constraints on the interaction between dark matter
  and baryons from cooling flow clusters}},
  \href{https://doi.org/10.1103/PhysRevLett.87.061301}{\emph{Phys. Rev. Lett.}
  {\bfseries 87} (2001) 061301}
  [\href{https://arxiv.org/abs/astro-ph/0106458}{{\ttfamily
  astro-ph/0106458}}].

\bibitem{Chuzhoy:2004bc}
L.~{Chuzhoy} and A.~{Nusser}, \emph{{Consequences of Short-Range Interactions
  between Dark Matter and Protons in Galaxy Clusters}},
  \href{https://doi.org/10.1086/504505}{\emph{\apj} {\bfseries 645} (2006) 950}
  [\href{https://arxiv.org/abs/astro-ph/0408184}{{\ttfamily
  astro-ph/0408184}}].

\bibitem{Hu:2007ai}
J.~Hu and Y.-Q.~Lou, \emph{{Collisional interaction limits between dark matters
  and baryons in `cooling flow' clusters}},
  \href{https://doi.org/10.1111/j.1365-2966.2007.12755.x}{\emph{Mon. Not. Roy.
  Astron. Soc.} {\bfseries 384} (2008) 814}
  [\href{https://arxiv.org/abs/0711.3555}{{\ttfamily 0711.3555}}].

\bibitem{PandaX:2022xqx}
{\scshape PandaX} collaboration, \emph{{Search for Light Dark Matter with
  Ionization Signals in the PandaX-4T Experiment}},
  \href{https://doi.org/10.1103/PhysRevLett.130.261001}{\emph{Phys. Rev. Lett.}
  {\bfseries 130} (2023) 261001}
  [\href{https://arxiv.org/abs/2212.10067}{{\ttfamily 2212.10067}}].

\bibitem{LZ:2022lsv}
{\scshape LZ} collaboration, \emph{{First Dark Matter Search Results from the
  LUX-ZEPLIN (LZ) Experiment}},
  \href{https://doi.org/10.1103/PhysRevLett.131.041002}{\emph{Phys. Rev. Lett.}
  {\bfseries 131} (2023) 041002}
  [\href{https://arxiv.org/abs/2207.03764}{{\ttfamily 2207.03764}}].

\bibitem{XENON:2023cxc}
{\scshape XENON} collaboration, \emph{{First Dark Matter Search with Nuclear
  Recoils from the XENONnT Experiment}},
  \href{https://doi.org/10.1103/PhysRevLett.131.041003}{\emph{Phys. Rev. Lett.}
  {\bfseries 131} (2023) 041003}
  [\href{https://arxiv.org/abs/2303.14729}{{\ttfamily 2303.14729}}].

\bibitem{Bringmann:2018cvk}
T.~Bringmann and M.~Pospelov, \emph{{Novel direct detection constraints on
  light dark matter}},
  \href{https://doi.org/10.1103/PhysRevLett.122.171801}{\emph{Phys. Rev. Lett.}
  {\bfseries 122} (2019) 171801}
  [\href{https://arxiv.org/abs/1810.10543}{{\ttfamily 1810.10543}}].

\bibitem{Bramante:2021dyx}
J.~Bramante, B.J.~Kavanagh and N.~Raj, \emph{{Scattering Searches for Dark
  Matter in Subhalos: Neutron Stars, Cosmic Rays, and Old Rocks}},
  \href{https://doi.org/10.1103/PhysRevLett.128.231801}{\emph{Phys. Rev. Lett.}
  {\bfseries 128} (2022) 231801}
  [\href{https://arxiv.org/abs/2109.04582}{{\ttfamily 2109.04582}}].

\bibitem{Bramante:2022pmn}
J.~Bramante, J.~Kumar, G.~Mohlabeng, N.~Raj and N.~Song, \emph{{Light dark
  matter accumulating in planets: Nuclear scattering}},
  \href{https://doi.org/10.1103/PhysRevD.108.063022}{\emph{Phys. Rev. D}
  {\bfseries 108} (2023) 063022}
  [\href{https://arxiv.org/abs/2210.01812}{{\ttfamily 2210.01812}}].

\bibitem{DiLuzio:2021qct}
L.~Di~Luzio et~al., \emph{{Probing the axion\textendash{}nucleon coupling with
  the next generation of~axion helioscopes}},
  \href{https://doi.org/10.1140/epjc/s10052-022-10061-1}{\emph{Eur. Phys. J. C}
  {\bfseries 82} (2022) 120}
  [\href{https://arxiv.org/abs/2111.06407}{{\ttfamily 2111.06407}}].

\bibitem{DiLuzio:2021ysg}
L.~Di~Luzio, M.~Fedele, M.~Giannotti, F.~Mescia and E.~Nardi, \emph{{Stellar
  evolution confronts axion models}},
  \href{https://doi.org/10.1088/1475-7516/2022/02/035}{\emph{JCAP} {\bfseries
  02} (2022) 035} [\href{https://arxiv.org/abs/2109.10368}{{\ttfamily
  2109.10368}}].

\bibitem{Carenza:2023wsm}
P.~Carenza, G.~Co, M.~Giannotti, A.~Lella, G.~Lucente, A.~Mirizzi et~al.,
  \emph{{Cross section~for supernova axion observation in neutrino water
  \v{C}herenkov detectors}},
  \href{https://doi.org/10.1103/PhysRevC.109.015501}{\emph{Phys. Rev. C}
  {\bfseries 109} (2024) 015501}
  [\href{https://arxiv.org/abs/2306.17055}{{\ttfamily 2306.17055}}].

\bibitem{Griest:1989wd}
K.~{Griest} and M.~{Kamionkowski}, \emph{{Unitarity limits on the mass and
  radius of dark-matter particles}},
  \href{https://doi.org/10.1103/PhysRevLett.64.615}{\emph{\prl} {\bfseries 64}
  (1990) 615}.

\bibitem{Cirelli:2024ssz}
M.~Cirelli, A.~Strumia and J.~Zupan, \emph{{Dark Matter}},
  \href{https://arxiv.org/abs/2406.01705}{{\ttfamily 2406.01705}}.

\bibitem{1984PhRvD..30..272W}
E.~{Witten}, \emph{{Cosmic separation of phases}},
  \href{https://doi.org/10.1103/PhysRevD.30.272}{\emph{\prd} {\bfseries 30}
  (1984) 272}.

\bibitem{PhysRevD.30.2379}
E.~Farhi and R.L.~Jaffe, \emph{Strange matter},
  \href{https://doi.org/10.1103/PhysRevD.30.2379}{\emph{Phys. Rev. D}
  {\bfseries 30} (1984) 2379}.

\bibitem{1984Natur.312..734D}
A.~{de Rujula} and S.L.~{Glashow}, \emph{{Nuclearites-a novel form of cosmic
  radiation}}, \href{https://doi.org/10.1038/312734a0}{\emph{\nat} {\bfseries
  312} (1984) 734}.

\bibitem{Zhitnitsky:2002qa}
A.R.~{Zhitnitsky}, \emph{{`Nonbaryonic' dark matter as baryonic colour
  superconductor}},
  \href{https://doi.org/10.1088/1475-7516/2003/10/010}{\emph{JCAP} {\bfseries
  10} (2003) 010} [\href{https://arxiv.org/abs/hep-ph/0202161}{{\ttfamily
  hep-ph/0202161}}].

\bibitem{Marsh:2015xka}
D.J.E.~{Marsh}, \emph{{Axion cosmology}},
  \href{https://doi.org/10.1016/j.physrep.2016.06.005}{\emph{\physrep}
  {\bfseries 643} (2016) 1} [\href{https://arxiv.org/abs/1510.07633}{{\ttfamily
  1510.07633}}].

\bibitem{Zhitnitsky:2021iwg}
A.~{Zhitnitsky}, \emph{{Axion quark nuggets. Dark matter and matter-antimatter
  asymmetry: Theory, observations and future experiments}},
  \href{https://doi.org/10.1142/S0217732321300172}{\emph{Modern Physics Letters
  A} {\bfseries 36} (2021) 2130017}
  [\href{https://arxiv.org/abs/2105.08719}{{\ttfamily 2105.08719}}].

\bibitem{Liang:2016tqc}
X.~{Liang} and A.~{Zhitnitsky}, \emph{{Axion field and the quark nugget's
  formation at the QCD phase transition}},
  \href{https://doi.org/10.1103/PhysRevD.94.083502}{\emph{Phys. Rev. D}
  {\bfseries 94} (2016) 083502}
  [\href{https://arxiv.org/abs/1606.00435}{{\ttfamily 1606.00435}}].

\bibitem{Ge:2017ttc}
S.~{Ge}, X.~{Liang} and A.~{Zhitnitsky}, \emph{{Cosmological C P -odd axion
  field as the coherent Berry's phase of the Universe}},
  \href{https://doi.org/10.1103/PhysRevD.96.063514}{\emph{Phys. Rev. D}
  {\bfseries 96} (2017) 063514}
  [\href{https://arxiv.org/abs/1702.04354}{{\ttfamily 1702.04354}}].

\bibitem{Ge:2017idw}
S.~{Ge}, X.~{Liang} and A.~{Zhitnitsky}, \emph{{Cosmological axion and a quark
  nugget dark matter model}},
  \href{https://doi.org/10.1103/PhysRevD.97.043008}{\emph{Phys. Rev. D}
  {\bfseries 97} (2018) 043008}
  [\href{https://arxiv.org/abs/1711.06271}{{\ttfamily 1711.06271}}].

\bibitem{Ge:2019voa}
S.~Ge, K.~Lawson and A.~Zhitnitsky, \emph{{The Axion Quark Nugget Dark Matter
  Model: Size Distribution and Survival Pattern}},
  \href{https://doi.org/10.1103/PhysRevD.99.116017}{\emph{Phys. Rev.}
  {\bfseries D99} (2019) 116017}
  [\href{https://arxiv.org/abs/1903.05090}{{\ttfamily 1903.05090}}].

\bibitem{Alford:2007xm}
M.G.~Alford, A.~Schmitt, K.~Rajagopal and T.~Schafer, \emph{{Color
  superconductivity in dense quark matter}},
  \href{https://doi.org/10.1103/RevModPhys.80.1455}{\emph{Rev. Mod. Phys.}
  {\bfseries 80} (2008) 1455}
  [\href{https://arxiv.org/abs/0709.4635}{{\ttfamily 0709.4635}}].

\bibitem{Raza:2018gpb}
N.~{Raza}, L.~{Van Waerbeke} and A.~{Zhitnitsky}, \emph{{Solar Corona Heating
  by the Axion Quark Nugget Dark Matter}},
  \href{https://doi.org/10.1103/PhysRevD.98.103527}{\emph{Phys. Rev. D}
  {\bfseries 98} (2018) 103527}
  [\href{https://arxiv.org/abs/1805.01897}{{\ttfamily 1805.01897}}].

\bibitem{Lawson:2019cvy}
K.~Lawson, X.~Liang, A.~Mead, M.S.R.~Siddiqui, L.~Van~Waerbeke and
  A.~Zhitnitsky, \emph{{Gravitationally trapped axions on Earth}},
  \href{https://doi.org/10.1103/PhysRevD.100.043531}{\emph{Phys. Rev.}
  {\bfseries D100} (2019) 043531}
  [\href{https://arxiv.org/abs/1905.00022}{{\ttfamily 1905.00022}}].

\bibitem{Gorham:2012hy}
P.~Gorham, \emph{{Antiquark nuggets as dark matter: New constraints and
  detection prospects}},
  \href{https://doi.org/10.1103/PhysRevD.86.123005}{\emph{Phys. Rev.}
  {\bfseries D86} (2012) 123005}
  [\href{https://arxiv.org/abs/1208.3697}{{\ttfamily 1208.3697}}].

\bibitem{Flambaum:2018ohm}
V.V.~Flambaum and A.R.~Zhitnitsky, \emph{{Primordial Lithium Puzzle and the
  Axion Quark Nugget Dark Matter Model}},
  \href{https://doi.org/10.1103/PhysRevD.99.023517}{\emph{Phys. Rev.}
  {\bfseries D99} (2019) 023517}
  [\href{https://arxiv.org/abs/1811.01965}{{\ttfamily 1811.01965}}].

\bibitem{Zhitnitsky:2006vt}
A.~Zhitnitsky, \emph{{Cold dark matter as compact composite objects}},
  \href{https://doi.org/10.1103/PhysRevD.74.043515}{\emph{Phys. Rev.}
  {\bfseries D74} (2006) 043515}
  [\href{https://arxiv.org/abs/astro-ph/0603064}{{\ttfamily
  astro-ph/0603064}}].

\bibitem{Lawson:2018qkc}
K.~Lawson and A.R.~Zhitnitsky, \emph{{The 21cm Absorption Line and Axion Quark
  Nugget Dark Matter Model}},
  \href{https://doi.org/10.1016/j.dark.2019.100295}{\emph{Phys. Dark Univ.}
  {\bfseries 24} (2019) 100295}
  [\href{https://arxiv.org/abs/1804.07340}{{\ttfamily 1804.07340}}].

\bibitem{Santillan:2020lbj}
O.P.~Santill\'an and A.~Morano, \emph{{Neutrino emission and initial evolution
  of axionic quark nuggets}},
  \href{https://doi.org/10.1103/PhysRevD.104.083530}{\emph{Phys. Rev. D}
  {\bfseries 104} (2021) 083530}
  [\href{https://arxiv.org/abs/2011.06747}{{\ttfamily 2011.06747}}].

\bibitem{SinghSidhu:2020cxw}
J.~Singh~Sidhu, R.J.~Scherrer and G.~Starkman, \emph{{Antimatter as Macroscopic
  Dark Matter}},
  \href{https://doi.org/10.1016/j.physletb.2020.135574}{\emph{Phys. Lett. B}
  {\bfseries 807} (2020) 135574}
  [\href{https://arxiv.org/abs/2006.01200}{{\ttfamily 2006.01200}}].

\bibitem{Forbes:2006ba}
M.M.~Forbes and A.R.~Zhitnitsky, \emph{{Diffuse x-rays: Directly observing dark
  matter?}}, \href{https://doi.org/10.1088/1475-7516/2008/01/023}{\emph{JCAP}
  {\bfseries 0801} (2008) 023}
  [\href{https://arxiv.org/abs/astro-ph/0611506}{{\ttfamily
  astro-ph/0611506}}].

\bibitem{Forbes:2008uf}
M.M.~Forbes and A.R.~Zhitnitsky, \emph{{WMAP Haze: Directly Observing Dark
  Matter?}}, \href{https://doi.org/10.1103/PhysRevD.78.083505}{\emph{Phys.
  Rev.} {\bfseries D78} (2008) 083505}
  [\href{https://arxiv.org/abs/0802.3830}{{\ttfamily 0802.3830}}].

\bibitem{Forbes:2009wg}
M.M.~Forbes, K.~Lawson and A.R.~Zhitnitsky, \emph{{The Electrosphere of
  Macroscopic 'Quark Nuclei': A Source for Diffuse MeV Emissions from Dark
  Matter}}, \href{https://doi.org/10.1103/PhysRevD.82.083510}{\emph{Phys. Rev.}
  {\bfseries D82} (2010) 083510}
  [\href{https://arxiv.org/abs/0910.4541}{{\ttfamily 0910.4541}}].

\bibitem{Zhitnitsky:2023znn}
A.~{Zhitnitsky}, \emph{{Structure formation paradigm and axion quark nugget
  dark matter model}},
  \href{https://doi.org/10.1016/j.dark.2023.101217}{\emph{Physics of the Dark
  Universe} {\bfseries 40} (2023) 101217}
  [\href{https://arxiv.org/abs/2302.00010}{{\ttfamily 2302.00010}}].

\bibitem{Ge:2020cho}
S.~{Ge}, H.~{Rachmat}, M.S.R.~{Siddiqui}, L.~{Van Waerbeke} and
  A.~{Zhitnitsky}, \emph{{X-ray annual modulation observed by XMM-Newton and
  Axion Quark Nugget dark matter}},
  \href{https://doi.org/10.1016/j.dark.2022.101031}{\emph{Physics of the Dark
  Universe} {\bfseries 36} (2022) 101031}
  [\href{https://arxiv.org/abs/2004.00632}{{\ttfamily 2004.00632}}].

\bibitem{Oaknin:2004mn}
D.H.~Oaknin and A.R.~Zhitnitsky, \emph{{511-KeV photons from color
  superconducting dark matter}},
  \href{https://doi.org/10.1103/PhysRevLett.94.101301}{\emph{Phys. Rev. Lett.}
  {\bfseries 94} (2005) 101301}
  [\href{https://arxiv.org/abs/hep-ph/0406146}{{\ttfamily hep-ph/0406146}}].

\bibitem{Zhitnitsky:2006tu}
A.~Zhitnitsky, \emph{{The Width of the 511-KeV Line from the Bulge of the
  Galaxy}}, \href{https://doi.org/10.1103/PhysRevD.76.103518}{\emph{Phys. Rev.}
  {\bfseries D76} (2007) 103518}
  [\href{https://arxiv.org/abs/astro-ph/0607361}{{\ttfamily
  astro-ph/0607361}}].

\bibitem{Lawson:2007kp}
K.~Lawson and A.R.~Zhitnitsky, \emph{{Diffuse cosmic gamma-rays at 1-20 MeV: A
  trace of the dark matter?}},
  \href{https://doi.org/10.1088/1475-7516/2008/01/022}{\emph{JCAP} {\bfseries
  0801} (2008) 022} [\href{https://arxiv.org/abs/0704.3064}{{\ttfamily
  0704.3064}}].

\bibitem{2016MNRAS.463.1797D}
K.~{Dolag}, E.~{Komatsu} and R.~{Sunyaev}, \emph{{SZ effects in the Magneticum
  Pathfinder simulation: comparison with the Planck, SPT, and ACT results}},
  \href{https://doi.org/10.1093/mnras/stw2035}{\emph{\mnras} {\bfseries 463}
  (2016) 1797} [\href{https://arxiv.org/abs/1509.05134}{{\ttfamily
  1509.05134}}].

\bibitem{2009ApJS..180..330K}
E.~{Komatsu}, J.~{Dunkley}, M.R.~{Nolta}, C.L.~{Bennett}, B.~{Gold},
  G.~{Hinshaw} et~al., \emph{{Five-Year Wilkinson Microwave Anisotropy Probe
  Observations: Cosmological Interpretation}},
  \href{https://doi.org/10.1088/0067-0049/180/2/330}{\emph{\apjs} {\bfseries
  180} (2009) 330} [\href{https://arxiv.org/abs/0803.0547}{{\ttfamily
  0803.0547}}].

\bibitem{2005MNRAS.364.1105S}
V.~{Springel}, \emph{{The cosmological simulation code GADGET-2}},
  \href{https://doi.org/10.1111/j.1365-2966.2005.09655.x}{\emph{\mnras}
  {\bfseries 364} (2005) 1105}
  [\href{https://arxiv.org/abs/astro-ph/0505010}{{\ttfamily
  astro-ph/0505010}}].

\bibitem{2016MNRAS.455.2110B}
A.M.~{Beck}, G.~{Murante}, A.~{Arth}, R.S.~{Remus}, A.F.~{Teklu},
  J.M.F.~{Donnert} et~al., \emph{{An improved SPH scheme for cosmological
  simulations}}, \href{https://doi.org/10.1093/mnras/stv2443}{\emph{\mnras}
  {\bfseries 455} (2016) 2110}
  [\href{https://arxiv.org/abs/1502.07358}{{\ttfamily 1502.07358}}].

\bibitem{2005MNRAS.364..753D}
K.~{Dolag}, F.~{Vazza}, G.~{Brunetti} and G.~{Tormen}, \emph{{Turbulent gas
  motions in galaxy cluster simulations: the role of smoothed particle
  hydrodynamics viscosity}},
  \href{https://doi.org/10.1111/j.1365-2966.2005.09630.x}{\emph{\mnras}
  {\bfseries 364} (2005) 753}
  [\href{https://arxiv.org/abs/astro-ph/0507480}{{\ttfamily
  astro-ph/0507480}}].

\bibitem{2009MNRAS.393...99W}
R.P.C.~{Wiersma}, J.~{Schaye} and B.D.~{Smith}, \emph{{The effect of
  photoionization on the cooling rates of enriched, astrophysical plasmas}},
  \href{https://doi.org/10.1111/j.1365-2966.2008.14191.x}{\emph{\mnras}
  {\bfseries 393} (2009) 99} [\href{https://arxiv.org/abs/0807.3748}{{\ttfamily
  0807.3748}}].

\bibitem{2005MNRAS.361..776S}
V.~{Springel}, T.~{Di Matteo} and L.~{Hernquist}, \emph{{Modelling feedback
  from stars and black holes in galaxy mergers}},
  \href{https://doi.org/10.1111/j.1365-2966.2005.09238.x}{\emph{\mnras}
  {\bfseries 361} (2005) 776}
  [\href{https://arxiv.org/abs/astro-ph/0411108}{{\ttfamily
  astro-ph/0411108}}].

\bibitem{2003MNRAS.339..289S}
V.~{Springel} and L.~{Hernquist}, \emph{{Cosmological smoothed particle
  hydrodynamics simulations: a hybrid multiphase model for star formation}},
  \href{https://doi.org/10.1046/j.1365-8711.2003.06206.x}{\emph{\mnras}
  {\bfseries 339} (2003) 289}
  [\href{https://arxiv.org/abs/astro-ph/0206393}{{\ttfamily
  astro-ph/0206393}}].

\bibitem{2001cghr.confE..64H}
F.~{Haardt} and P.~{Madau}, \emph{{Modelling the UV/X-ray cosmic background
  with CUBA}},  in \emph{Clusters of Galaxies and the High Redshift Universe
  Observed in X-rays}, D.M.~{Neumann} and J.T.V.~{Tran}, eds., p.~64, Jan.,
  2001 [\href{https://arxiv.org/abs/astro-ph/0106018}{{\ttfamily
  astro-ph/0106018}}].

\bibitem{2007MNRAS.382.1050T}
L.~{Tornatore}, S.~{Borgani}, K.~{Dolag} and F.~{Matteucci}, \emph{{Chemical
  enrichment of galaxy clusters from hydrodynamical simulations}},
  \href{https://doi.org/10.1111/j.1365-2966.2007.12070.x}{\emph{\mnras}
  {\bfseries 382} (2007) 1050}
  [\href{https://arxiv.org/abs/0705.1921}{{\ttfamily 0705.1921}}].

\bibitem{2004MNRAS.349L..19T}
L.~{Tornatore}, S.~{Borgani}, F.~{Matteucci}, S.~{Recchi} and P.~{Tozzi},
  \emph{{Simulating the metal enrichment of the intracluster medium}},
  \href{https://doi.org/10.1111/j.1365-2966.2004.07689.x}{\emph{\mnras}
  {\bfseries 349} (2004) L19}
  [\href{https://arxiv.org/abs/astro-ph/0401576}{{\ttfamily
  astro-ph/0401576}}].

\bibitem{2014arXiv1412.6533A}
A.~{Arth}, K.~{Dolag}, A.M.~{Beck}, M.~{Petkova} and H.~{Lesch},
  \emph{{Anisotropic thermal conduction in galaxy clusters with MHD in
  Gadget}}, {\emph{arXiv e-prints} (2014) arXiv:1412.6533}
  [\href{https://arxiv.org/abs/1412.6533}{{\ttfamily 1412.6533}}].

\bibitem{2010MNRAS.401.1670F}
D.~{Fabjan}, S.~{Borgani}, L.~{Tornatore}, A.~{Saro}, G.~{Murante} and
  K.~{Dolag}, \emph{{Simulating the effect of active galactic nuclei feedback
  on the metal enrichment of galaxy clusters}},
  \href{https://doi.org/10.1111/j.1365-2966.2009.15794.x}{\emph{\mnras}
  {\bfseries 401} (2010) 1670}
  [\href{https://arxiv.org/abs/0909.0664}{{\ttfamily 0909.0664}}].

\bibitem{2014MNRAS.442.2304H}
M.~{Hirschmann}, K.~{Dolag}, A.~{Saro}, L.~{Bachmann}, S.~{Borgani} and
  A.~{Burkert}, \emph{{Cosmological simulations of black hole growth: AGN
  luminosities and downsizing}},
  \href{https://doi.org/10.1093/mnras/stu1023}{\emph{\mnras} {\bfseries 442}
  (2014) 2304} [\href{https://arxiv.org/abs/1308.0333}{{\ttfamily 1308.0333}}].

\bibitem{2018ApSpe..72..663H}
R.~{Hill}, K.W.~{Masui} and D.~{Scott}, \emph{{The Spectrum of the Universe}},
  \href{https://doi.org/10.1177/0003702818767133}{\emph{Applied Spectroscopy}
  {\bfseries 72} (2018) 663}
  [\href{https://arxiv.org/abs/1802.03694}{{\ttfamily 1802.03694}}].

\bibitem{2011JCAP...07..025K}
A.~{Kogut}, D.J.~{Fixsen}, D.T.~{Chuss}, J.~{Dotson}, E.~{Dwek}, M.~{Halpern}
  et~al., \emph{{The Primordial Inflation Explorer (PIXIE): a nulling
  polarimeter for cosmic microwave background observations}},
  \href{https://doi.org/10.1088/1475-7516/2011/07/025}{\emph{\jcap} {\bfseries
  2011} (2011) 025} [\href{https://arxiv.org/abs/1105.2044}{{\ttfamily
  1105.2044}}].

\bibitem{2013arXiv1306.2259P}
{PRISM Collaboration}, P.~{Andre}, C.~{Baccigalupi}, D.~{Barbosa},
  J.~{Bartlett}, N.~{Bartolo} et~al., \emph{{PRISM (Polarized Radiation Imaging
  and Spectroscopy Mission): A White Paper on the Ultimate Polarimetric
  Spectro-Imaging of the Microwave and Far-Infrared Sky}},
  \href{https://doi.org/10.48550/arXiv.1306.2259}{\emph{arXiv e-prints} (2013)
  arXiv:1306.2259} [\href{https://arxiv.org/abs/1306.2259}{{\ttfamily
  1306.2259}}].

\bibitem{2014JCAP...02..006A}
P.~{Andr{\'e}}, C.~{Baccigalupi}, A.~{Banday}, D.~{Barbosa}, B.~{Barreiro},
  J.~{Bartlett} et~al., \emph{{PRISM (Polarized Radiation Imaging and
  Spectroscopy Mission): an extended white paper}},
  \href{https://doi.org/10.1088/1475-7516/2014/02/006}{\emph{\jcap} {\bfseries
  2014} (2014) 006} [\href{https://arxiv.org/abs/1310.1554}{{\ttfamily
  1310.1554}}].

\bibitem{2019BAAS...51g.113K}
A.~{Kogut}, M.H.~{Abitbol}, J.~{Chluba}, J.~{Delabrouille}, D.~{Fixsen},
  J.C.~{Hill} et~al., \emph{{CMB Spectral Distortions: Status and Prospects}},
  in \emph{Bulletin of the American Astronomical Society}, vol.~51, p.~113,
  Sept., 2019, \href{https://doi.org/10.48550/arXiv.1907.13195}{DOI}
  [\href{https://arxiv.org/abs/1907.13195}{{\ttfamily 1907.13195}}].

\bibitem{2021ExA....51.1515C}
J.~{Chluba}, M.H.~{Abitbol}, N.~{Aghanim}, Y.~{Ali-Ha{\"\i}moud}, M.~{Alvarez},
  K.~{Basu} et~al., \emph{{New horizons in cosmology with spectral distortions
  of the cosmic microwave background}},
  \href{https://doi.org/10.1007/s10686-021-09729-5}{\emph{Experimental
  Astronomy} {\bfseries 51} (2021) 1515}
  [\href{https://arxiv.org/abs/1909.01593}{{\ttfamily 1909.01593}}].

\bibitem{2019MNRAS.484.4127A}
D.~{Alonso}, J.~{Sanchez}, A.~{Slosar} and {LSST Dark Energy Science
  Collaboration}, \emph{{A unified pseudo-C$_{{\ensuremath{\ell}}}$
  framework}}, \href{https://doi.org/10.1093/mnras/stz093}{\emph{\mnras}
  {\bfseries 484} (2019) 4127}
  [\href{https://arxiv.org/abs/1809.09603}{{\ttfamily 1809.09603}}].

\bibitem{2012ApJ...755...70R}
C.L.~{Reichardt}, L.~{Shaw}, O.~{Zahn}, K.A.~{Aird}, B.A.~{Benson},
  L.E.~{Bleem} et~al., \emph{{A Measurement of Secondary Cosmic Microwave
  Background Anisotropies with Two Years of South Pole Telescope
  Observations}}, \href{https://doi.org/10.1088/0004-637X/755/1/70}{\emph{\apj}
  {\bfseries 755} (2012) 70} [\href{https://arxiv.org/abs/1111.0932}{{\ttfamily
  1111.0932}}].

\bibitem{2019ConPh..60...23M}
K.~{Mattila} and P.~{V{\"a}is{\"a}nen}, \emph{{Extragalactic background light:
  inventory of light throughout the cosmic history}},
  \href{https://doi.org/10.1080/00107514.2019.1586130}{\emph{Contemporary
  Physics} {\bfseries 60} (2019) 23}
  [\href{https://arxiv.org/abs/1905.08825}{{\ttfamily 1905.08825}}].

\bibitem{2022ApJ...927L...8L}
T.R.~{Lauer}, M.~{Postman}, J.R.~{Spencer}, H.A.~{Weaver}, S.A.~{Stern},
  G.R.~{Gladstone} et~al., \emph{{Anomalous Flux in the Cosmic Optical
  Background Detected with New Horizons Observations}},
  \href{https://doi.org/10.3847/2041-8213/ac573d}{\emph{\apjl} {\bfseries 927}
  (2022) L8} [\href{https://arxiv.org/abs/2202.04273}{{\ttfamily 2202.04273}}].

\bibitem{2023ApJ...945...45S}
T.~{Symons}, M.~{Zemcov}, A.~{Cooray}, C.~{Lisse} and A.R.~{Poppe}, \emph{{A
  Measurement of the Cosmic Optical Background and Diffuse Galactic Light
  Scaling from the R < 50 au New Horizons-LORRI Data}},
  \href{https://doi.org/10.3847/1538-4357/acaa37}{\emph{\apj} {\bfseries 945}
  (2023) 45} [\href{https://arxiv.org/abs/2212.07449}{{\ttfamily 2212.07449}}].

\bibitem{2022PhRvL.129w1301B}
J.L.~{Bernal}, G.~{Sato-Polito} and M.~{Kamionkowski}, \emph{{Cosmic Optical
  Background Excess, Dark Matter, and Line-Intensity Mapping}},
  \href{https://doi.org/10.1103/PhysRevLett.129.231301}{\emph{\prl} {\bfseries
  129} (2022) 231301} [\href{https://arxiv.org/abs/2203.11236}{{\ttfamily
  2203.11236}}].

\bibitem{2022PhRvD.106j3505N}
K.~{Nakayama} and W.~{Yin}, \emph{{Anisotropic cosmic optical background bound
  for decaying dark matter in light of the LORRI anomaly}},
  \href{https://doi.org/10.1103/PhysRevD.106.103505}{\emph{\prd} {\bfseries
  106} (2022) 103505} [\href{https://arxiv.org/abs/2205.01079}{{\ttfamily
  2205.01079}}].

\bibitem{2015NatCo...6.7945M}
K.~{Mitchell-Wynne}, A.~{Cooray}, Y.~{Gong}, M.~{Ashby}, T.~{Dolch},
  H.~{Ferguson} et~al., \emph{{Ultraviolet luminosity density of the universe
  during the epoch of reionization}},
  \href{https://doi.org/10.1038/ncomms8945}{\emph{Nature Communications}
  {\bfseries 6} (2015) 7945}
  [\href{https://arxiv.org/abs/1509.02935}{{\ttfamily 1509.02935}}].

\bibitem{2024arXiv240513496C}
J.C.~{Cuillandre}, E.~{Bertin}, M.~{Bolzonella}, H.~{Bouy}, S.~{Gwyn},
  S.~{Isani} et~al., \emph{{Euclid: Early Release Observations -- Programme
  overview and pipeline for compact- and diffuse-emission photometry}},
  \href{https://doi.org/10.48550/arXiv.2405.13496}{\emph{arXiv e-prints} (2024)
  arXiv:2405.13496} [\href{https://arxiv.org/abs/2405.13496}{{\ttfamily
  2405.13496}}].

\bibitem{2024MNRAS.528.5019H}
R.~{Hill}, D.~{Scott}, D.J.~{McLeod}, R.J.~{McLure}, S.C.~{Chapman} and
  J.S.~{Dunlop}, \emph{{An optimal ALMA image of the Hubble Ultra Deep Field in
  the era of JWST: obscured star formation and the cosmic far-infrared
  background}}, \href{https://doi.org/10.1093/mnras/stae346}{\emph{\mnras}
  {\bfseries 528} (2024) 5019}
  [\href{https://arxiv.org/abs/2309.10988}{{\ttfamily 2309.10988}}].

\bibitem{2020PASP..132c5001L}
M.~{Lacy}, S.A.~{Baum}, C.J.~{Chandler}, S.~{Chatterjee}, T.E.~{Clarke},
  S.~{Deustua} et~al., \emph{{The Karl G. Jansky Very Large Array Sky Survey
  (VLASS). Science Case and Survey Design}},
  \href{https://doi.org/10.1088/1538-3873/ab63eb}{\emph{\pasp} {\bfseries 132}
  (2020) 035001} [\href{https://arxiv.org/abs/1907.01981}{{\ttfamily
  1907.01981}}].

\bibitem{2023MNRAS.525.1443L}
S.~{Lim}, R.~{Hill}, D.~{Scott}, L.~{van Waerbeke}, J.-C.~{Cuillandre},
  R.G.~{Carlberg} et~al., \emph{{Constraints on galaxy formation from the
  cosmic-far-infrared-background - optical-imaging cross-correlation using
  Herschel and UNIONS}},
  \href{https://doi.org/10.1093/mnras/stad2177}{\emph{\mnras} {\bfseries 525}
  (2023) 1443} [\href{https://arxiv.org/abs/2203.16545}{{\ttfamily
  2203.16545}}].

\bibitem{2024arXiv240617946S}
J.S.~{Sommer}, K.~{Dolag}, L.M.~{B{\"o}ss}, I.~{Khabibullin}, X.~{Liang},
  L.~{Van Waerbeke} et~al., \emph{{The Glow of Axion Quark Nugget Dark Matter:
  (II) Galaxy Clusters}},
  \href{https://doi.org/10.48550/arXiv.2406.17946}{\emph{arXiv e-prints} (2024)
  arXiv:2406.17946} [\href{https://arxiv.org/abs/2406.17946}{{\ttfamily
  2406.17946}}].

\bibitem{sekatchev2024prep}
M.~Sekatchev and et~al., \emph{Axion quark nuggets: A recipe for a glowing
  milky way?}, {\emph{In preparation} (2024) }.

\bibitem{majidi2024prep}
F.~Majidi and et~al., \emph{Axion quark nuggets in the pre-recombination era},
  {\emph{In preparation} (2024) }.

\bibitem{2019JCAP...04..025G}
D.~{Green}, P.D.~{Meerburg} and J.~{Meyers}, \emph{{Aspects of dark matter
  annihilation in cosmology}},
  \href{https://doi.org/10.1088/1475-7516/2019/04/025}{\emph{\jcap} {\bfseries
  2019} (2019) 025} [\href{https://arxiv.org/abs/1804.01055}{{\ttfamily
  1804.01055}}].

\bibitem{2020JCAP...07..020S}
K.~{Short}, J.L.~{Bernal}, A.~{Raccanelli}, L.~{Verde} and J.~{Chluba},
  \emph{{Enlightening the dark ages with dark matter}},
  \href{https://doi.org/10.1088/1475-7516/2020/07/020}{\emph{\jcap} {\bfseries
  2020} (2020) 020} [\href{https://arxiv.org/abs/1912.07409}{{\ttfamily
  1912.07409}}].

\bibitem{2011ApJ...734....5F}
D.J.~{Fixsen}, A.~{Kogut}, S.~{Levin}, M.~{Limon}, P.~{Lubin}, P.~{Mirel}
  et~al., \emph{{ARCADE 2 Measurement of the Absolute Sky Brightness at 3-90
  GHz}}, \href{https://doi.org/10.1088/0004-637X/734/1/5}{\emph{\apj}
  {\bfseries 734} (2011) 5} [\href{https://arxiv.org/abs/0901.0555}{{\ttfamily
  0901.0555}}].

\bibitem{2020Natur.585..357H}
C.R.~{Harris}, K.J.~{Millman}, S.J.~{van der Walt}, R.~{Gommers},
  P.~{Virtanen}, D.~{Cournapeau} et~al., \emph{{Array programming with NumPy}},
  \href{https://doi.org/10.1038/s41586-020-2649-2}{\emph{\nat} {\bfseries 585}
  (2020) 357} [\href{https://arxiv.org/abs/2006.10256}{{\ttfamily
  2006.10256}}].

\bibitem{2020NatMe..17..261V}
P.~{Virtanen}, R.~{Gommers}, T.E.~{Oliphant}, M.~{Haberland}, T.~{Reddy},
  D.~{Cournapeau} et~al., \emph{{SciPy 1.0: fundamental algorithms for
  scientific computing in Python}},
  \href{https://doi.org/10.1038/s41592-019-0686-2}{\emph{Nature Methods}
  {\bfseries 17} (2020) 261}
  [\href{https://arxiv.org/abs/1907.10121}{{\ttfamily 1907.10121}}].

\bibitem{2018AJ....156..123A}
{Astropy Collaboration}, A.M.~{Price-Whelan}, B.M.~{Sip{\H{o}}cz},
  H.M.~{G{\"u}nther}, P.L.~{Lim}, S.M.~{Crawford} et~al., \emph{{The Astropy
  Project: Building an Open-science Project and Status of the v2.0 Core
  Package}}, \href{https://doi.org/10.3847/1538-3881/aabc4f}{\emph{\aj}
  {\bfseries 156} (2018) 123}
  [\href{https://arxiv.org/abs/1801.02634}{{\ttfamily 1801.02634}}].

\bibitem{2007CSE.....9...90H}
J.D.~{Hunter}, \emph{{Matplotlib: A 2D Graphics Environment}},
  \href{https://doi.org/10.1109/MCSE.2007.55}{\emph{Computing in Science and
  Engineering} {\bfseries 9} (2007) 90}.

\end{thebibliography}\endgroup

%% or
%% [B] Manual formatting (see below)
%% (i) We suggest to always provide author, title and journal data or doi:
%% in short all the informations that clearly identify a document.
%% (ii) please avoid comments such as "For a review'', "For some examples",
%% "and references therein" or move them in the text. In general, please leave only references in the bibliography and move all
%% accessory text in footnotes.
%% (iii) Also, please have only one work for each \bibitem.

% \begin{thebibliography}{99}

% \bibitem{a}
% Author,
% \emph{Title},
% \emph{J. Abbrev.} {\bf vol} (year) pg.

% \bibitem{b}
% Author,
% \emph{Title},
% arxiv:1234.5678.

% \bibitem{c}
% Author,
% \emph{Title},
% Publisher (year).

% \end{thebibliography}
\end{document}